\documentclass[12pt]{article}
\usepackage{geometry} 
\usepackage{amsfonts}
\usepackage{amsmath}
\usepackage{graphicx}
\def\hyph{-\penalty0\hskip0pt\relax} 

\title{}
\author{}

\begin{document}

\title{Non-intersecting Brownian walkers and Yang-Mills theory on the sphere}

\author{Peter J. Forrester$^1$, Satya N. Majumdar$^2$, Gr\'egory Schehr$^3$ 
\\ $^1$ \small{\textit{Department of Mathematics and Statistics}}, \\
\small{\textit{The University of Melbourne, Victoria 3010, Australia}}\\
\noindent $^2$ \small{\textit{Univ. Paris Sud, CNRS, LPTMS,
UMR 8626, Orsay F-91405, France}}\\
\noindent $^3$  \small{\textit{Univ. Paris Sud, CNRS, LPT,
UMR 8627, Orsay F-91405, France}} }

\maketitle

\begin{abstract} 

\noindent We study a system of $N$ non-intersecting Brownian motions
on a line segment $[0,L]$ with periodic, absorbing and reflecting
boundary conditions. We show that the normalized reunion probabilities  
of these Brownian motions in the three models can be mapped to
the partition function of two-dimensional continuum Yang-Mills theory
on a sphere respectively with gauge groups ${\rm U}(N)$, ${\rm Sp}(2N)$ and
${\rm SO}(2N)$. Consequently, we show that in each of these Brownian
motion models, as one varies the system size $L$, a third order phase 
transition occurs at a critical value $L=L_c(N)\sim \sqrt{N}$
in the large $N$ limit. Close to the critical point, the reunion probability,
properly centered and scaled, is identical to the Tracy-Widom distribution
describing the probability distribution of the largest eigenvalue of
a random matrix. For the periodic case we obtain the Tracy-Widom
distribution corresponding to the GUE random matrices, while for the absorbing
and reflecting cases we get the Tracy-Widom distribution corresponding 
to GOE random matrices. In the absorbing case, the reunion probability
is also identified as the maximal height of $N$    
non-intersecting Brownian excursions (``watermelons" with a wall)
whose distribution in the asymptotic scaling limit is then
described by GOE Tracy-Widom law.
In addition, large deviation formulas for 
the maximum height are also computed.

\end{abstract}

\section{Introduction}

\subsection{Background results} 

It is a well-known result that ${\rm U}(N)$ lattice QCD in two dimensions with 
Wilson's action~\cite{Wilson74} exhibits a third order phase transition in the 
large $N$ limit 
\cite{GW80,Wadia80}. This is shown by forming the partition function for the 
plaquettes, 
which factorizes as a product of partition functions for each individual 
plaquette. The latter is identified with a zero-dimensional unitary matrix model 
having partition function given by
\begin{equation}\label{firsteqn}
G_N(b):=\Big \langle e^{bN{\rm Tr}(U+U^\dagger)} \Big \rangle_{U \in {\rm U}(N)},
\end{equation} 
where the matrices $U\in {\rm U}(N)$ are chosen with Haar measure and $b$ is the 
scaled coupling.

The matrix integral (\ref{firsteqn}) depends only on the $N$ eigenvalues of $U$, 
and in terms of these variables it can be written 
\begin{equation}\label{secondeqn} 
G_N(b)=\frac{1}{(2\pi)^NN!}\int_0^{2\pi}d\theta_1\cdots \int_0^{2\pi}d\theta_N 
\prod_{l=1}^Ne^{2bN\cos\theta_l} \prod_{1\leq j<k\leq N} |e^{i\theta_k}-e^{i 
\theta_j}|^2. 
\end{equation} 
This can be interpreted as a partition function for 
a classical gas of charged particles, confined to the unit circle, and repelling 
via logarithmic pair potential $-(1/2)\log|e^{i\theta}-e^{i \phi}|$ at the 
inverse 
temperature $\beta=2$. The charges are also subject to the extensive one-body 
potential $bN\cos\theta$. In the form (\ref{secondeqn}) the $N\rightarrow 
\infty$ limit can be computed with the result \cite{GW80}
\begin{equation}\label{secondeqn2}
\lim_{N\rightarrow \infty} \frac{1}{N^2}\log G_N(b)=\left\{ 
\begin{tabular}{ll}
$b^2,$ & $0<b<\frac{1}{2}$  \\
$2b-\frac{3}{4}-\frac{1}{2}\log2b$, & $b>\frac{1}{2}$,
\end{tabular}
\right. 
\end{equation}
which is indeed discontinuous in the third derivative at $b=1/2$. 

\begin{figure}[ht]
\begin{center}
\includegraphics[width=10cm]{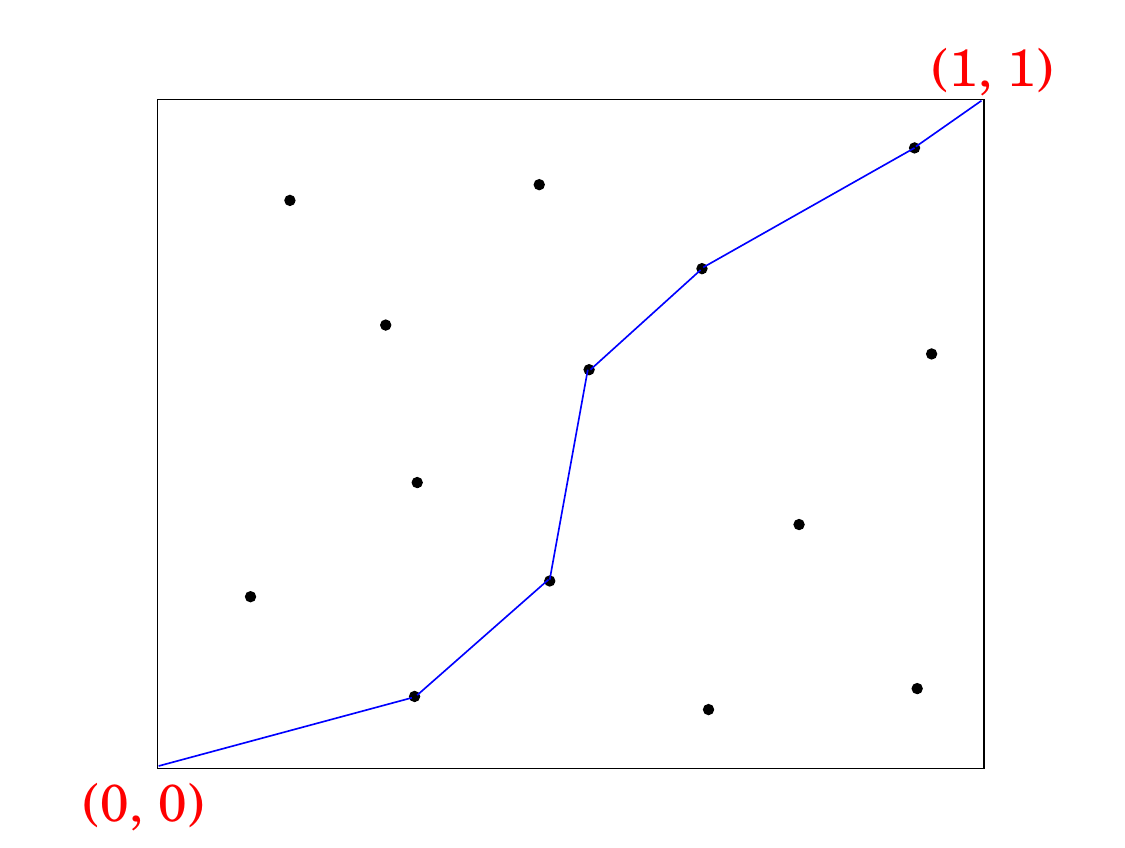}
\caption{Points (black dots) are distributed uniformly in a unit square
with mean density $\lambda^2$. The length of any up-right path connecting
the origin $(0,0)$ and the corner $(1,1)$ is defined by the number of
points on the line. A path with the longest length is shown by the (blue) solid line.}\label{fig:hammersley}
\end{center}
\end{figure}

Some fifteen years after the work \cite{GW80,Wadia80} the same matrix 
integral 
(\ref{firsteqn}) appeared in a completely different setting. Consider a unit 
square, and place points uniformly at random, with the number of points 
$n$ chosen 
according to the Poisson distribution 
$P(n)=\frac{\lambda^{2n}}{n!}e^{-\lambda^2}$ with mean $\langle 
n\rangle=\lambda^2$. 
Starting at $(0,0)$ and finishing at $(1,1)$ form a piecewise linear path by 
joining dots with line segments of positive slope. 
There are evidently many such paths. From them, choose a
{\em longest} path, {\it i.e.} a path with the maximum number
of dots on it (see Fig. (\ref{fig:hammersley})). 
Let $h^\square$ denote the length of this longest path. 
Clearly $h^\square$ is a random variable that fluctuates from
one configuration of points to another and its probability
distribution has been studied extensively in the context  
of the directed last passage percolation model due to Hammersley (see 
e.g. \cite[\S 10.9]{Fo10}). For the cumulative distribution of
$h^\square$, one gets 
\cite{Ge90,Ra98} (the first of the 
references gives a Toeplitz integral form equivalent to (\ref{secondeqn}), while 
the second identifies the matrix integral explicitly)
\begin{equation}
{\rm Pr}(h^\square<N)=e^{-\lambda^2} \Big \langle e^{ \lambda {\rm Tr}(U+U^\dagger)}
\Big \rangle_{U \in {\rm U}(N)}.
\label{cum-hamm}
\end{equation}
In the limit of large $\lambda$, we set $\lambda= bN$,
it follows from (\ref{secondeqn}) that in the large $N$ limit
\begin{equation}
{\rm Pr}(h^\square<N)=\left\{ 
\begin{tabular}{ll}
$1,$ & $0<b<\frac{1}{2}$  \\
$e^{N^2(-b^2+2b-3/4-(1/2)\log2b)}$, & $b>\frac{1}{2}$,
\end{tabular}
\right. 
\end{equation}
hence providing for (\ref{secondeqn2}) the interpretation as a large deviation 
formula for a probabilistic quantity in a statistical model \cite{Jo98a}.

The desire to relate 2d lattice QCD to string theory focussed attention on the 
so-called double scaling limit of matrix integrals. Here, in addition to 
$N\rightarrow \infty$ the coupling is tuned in the neighbourhood of the critical 
point $b=1/2$ to give a well-defined scaling limit. 
It turns out 
that if one zooms in the neighbourhood of the critical point $b=1/2$
and magnifies it by a factor $N^{2/3}$, i.e., one takes
the limit $(1/2-b)\to 0$, $N\to \infty$, but keeping the product
$t=2^{4/3}(1/2-b)N^{2/3}$ fixed, then the second derivative
of the free energy (the specific heat) tends to a function
of the single scaled variable $t$.
In the case of the matrix 
integral 
(\ref{firsteqn}) this double scaling limit was first analyzed by Periwal and 
Shevitz \cite{PS90a} whose result, using our notations, can be translated in the 
following form:
setting $b=1/2- 2^{-4/3} N^{-2/3}\, t$
\begin{equation}\label{fortheqn}
\frac{d^2}{dt^2}\lim_{N \rightarrow \infty}\left[\log G_N (b) - b^2 
N^2\right] = -q^2(t),
\end{equation}
where $q(t)$ satisfies the special case $\alpha=0$ of the Painlev\'{e} 
\textrm{II} differential equation
\begin{equation} \label{forthequation2} u''=2u^3+tu + \alpha.
\end{equation}
In other words
\begin{equation}
q''(t) = 2q^3(t) + t q(t).
\label{qdiff}
\end{equation} 
However no boundary condition was specified until Gross and Matytsin \cite{GM94} 
refined the working of \cite{PS90a} to obtain a result which implies
\begin{equation}\label{bc}
q(t)\mathop{\sim}\limits_{t\rightarrow \infty} {\rm Ai}(t)
\end{equation}
(see also the earlier reference \cite{MP90} for identification of Ai$(t)$ in a 
similar context, and Section \ref{sdd} below), where ${\rm Ai}(t)$ denotes the 
Airy function.

In the context of the Hammersley model, recalling that in (\ref{cum-hamm}) $\lambda = b N$, we see that as a consequence of (\ref{fortheqn})
\begin{equation} \label{fiftheqn}
\lim_{N\rightarrow \infty} {\rm Pr} 
\bigg(\frac{h^\square-N}{(N/2)^{1/3}}<t\bigg)= 
\exp\bigg(-\int_t^\infty (s-t)q^2(s)ds\bigg). \end{equation}
This result was first obtained in the probability literature \cite{BDJ98} 
independent of the working of 
\cite{PS90a}. In fact, the authors of \cite{BDJ98} were interested
in studying a `de-Poissonized' version of the Hammersley model, where
the number of dots in the unit square is a fixed number $N$, and not 
a Poisson distributed random variable. This `de-Poissonized' Hammersley
model is, in turn, related to the so-called Ulam problem
where one studies the statistics of the length of the longest
increasing subsequence of a random permutation of $N$ integers (for
a survey see ~\cite{AD99,Ma07}).
The length of the longest path in the `de-Poissonized' Hammersley
model has the same probability distribution as the length
of the longest increasing subsequence in the random permutation. 
The Ulam problem has recently turned out to be a key model
that connects various problems in combinatorics, physics
and probability~\cite{AD99,PS99,Ma07}.

Remarkably, the right-hand side (rhs) of (\ref{fiftheqn}) admits a second interpretation within 
random matrix theory. Consider the Gaussian unitary ensemble (GUE)
of the set of $N\times N$ 
complex Hermitian matrices $X$ with measure proportional to $e^{-{\rm Tr X}^2}$.
Let $\lambda_{\rm max}$ denote the largest eigenvalue.
Its average, in the large $N$ limit, is
simply $\langle \lambda_{\rm max}\rangle \simeq \sqrt{2N}$.
The typical fluctuations of $\lambda_{\rm max}$ around
its average are very small of order $\sim N^{-1/6}$.
It turns out that the probability distribution of 
these typical fluctuations have a limiting distribution
\cite{TW94a}
\begin{equation} \label{F2}
\lim_{N\rightarrow \infty} {\rm Pr}\bigg({2^{1/2}N^{1/6}} (\lambda_{\rm 
max}-\sqrt{2N})<t \bigg)
=\exp\bigg( - \int_t^\infty (s-t)q^2(s)ds \bigg):= \mathcal{F}_2(t),
\end{equation}
known as the $\beta=2$ Tracy-Widom distribution. This distribution
function has recently appeared in a number of problems ranging from
physics to biology~\cite{PS99,Ma07}. 
So (\ref{fiftheqn}) and (\ref{F2}) link distribution 
functions in two seemingly 
unrelated problems. In addition, the same Tracy-Widom function also
appears in the scaled specific heat of the ${\rm U}(N)$ lattice QCD 
in two dimensions as demonstrated by Eq. (\ref{fortheqn}).  

\subsection{Statement of problems and summary of new results} 

In the previous subsection, we have seen that the partition function
of a two-dimensional field theory model [${\rm U}(N)$ lattice QCD with Wilson 
action], when multiplied
by a factor $e^{-\lambda^2}$ (see Eq. (\ref{cum-hamm})), can be 
interpreted as the cumulative probability distribution of
a certain random variable in the statistical physics/probability problem of
Hammersley's directed last percolation model. It is then a natural
question to ask if such connections can be established between
other field theory models (on one side) and statistical physics models (on 
the other side).
In this paper we establish another connection, namely between
the continuum Yang-Mills gauge theory in two dimensions on a sphere
(the field theory model) and the system of $N$ non-intersecting
Brownian motions on a line segment $[0,L]$ (the statistical physics
model).  

Non-intersecting random walkers, first introduced
by de Gennes~\cite{deG68}, followed by Fisher~\cite{Fisher84}, 
have been studied extensively
in statistical physics as they appear in a variety of physical contexts 
ranging from wetting and melting all the way to polymers and vortex lines
in superconductors. Lattice versions of such walkers have
also beautiful combinatorial properties~\cite{KGV2000}.
Non-intersecting Brownian motions, defined in continuous space
and time, have also recently appeared in a number of contexts, in
particular its connection to the random matrix theory
have been noted in a variety of situations
~\cite{FP2006,Jo03,
KT2004,NM09,RS10,SMCR08,TW2007}. In this paper we introduce three
new models of non-intersecting Brownian motions and establish
their close connections to the Yang-Mills gauge theory in two dimensions on
a sphere.
  
Specifically, we consider a set of $N$ non-intersecting Brownian 
motions on a finite segment $[0,L]$ of the real line with
different boundary conditions.
Assuming that all the walkers start from the vicinity of 
the origin, we then
define the reunion probability as the probability that the walkers
reunite at the origin after a fixed interval of time which 
can be set to unity without any loss of generality. 
Next we `normalize' this reunion probability in a precise way to
be defined shortly. 
In one case, namely when both boundaries at $0$ and $L$ are absorbing,
one can relate this `normalized' reunion probability to the probability 
distribution
of the maximal height of $N$ non-intersecting Brownian excursions.
We show in this paper how to map this normalized reunion probability
in the Brownian motion models
to the exactly solvable partition function (up to a
multiplicative factor) of two-dimensional 
Yang-Mills theory on a sphere. The boundary conditions at the edges $0$
and $L$ select the gauge group of the associated Yang-Mills theory.
We consider three different boundary conditions: periodic (model I), 
absorbing (model II) and reflecting (model III) which correspond respectively
to the following gauge groups in the Yang-Mills theory: (I) periodic $\to$ 
${\rm U}(N)$ (II) absorbing $\to$ 
${\rm Sp}(2N)$ and (III) reflecting $\to$ ${\rm SO}(2N)$.  

Using the known results on the partition function from the Yang-Mills 
theory, we will show
how these normalized reunion probabilities in the Brownian motion models
can be related to the
limiting Tracy-Widom distribution of the largest eigenvalue in 
some particular random matrix ensembles.
The latter, in addition to the distribution $\mathcal{F}_2(t)$ 
relating to complex Hermitian matrices, involves its companion 
$\mathcal{F}_1(t)$ for real symmetric matrices. Explicitly with the GOE 
specified as the set of $N \times N$ real symmetric matrices $X$ with measure 
proportional to $e^{-{\rm Tr }X^2/2}$, and $\lambda_{\rm max}$ denoting the 
largest eigenvalue, one has \cite{TW96}
\begin{eqnarray}\label{F2-GOE}
\lim_{N\rightarrow \infty} 
{\rm Pr} \bigg(\sqrt{2}N^{\frac{1}{6}}(\lambda_{\rm max}-\sqrt{2N})<t\bigg)&=&
\exp\bigg( -\frac{1}{2} \int_t^\infty \left( 
\left(s-t\right)q^2(s)-q(s)\right)\,ds 
\bigg) \nonumber \\ 
&\mathrel{\mathop:}=& \mathcal{F}_1(t) \;.
\end{eqnarray}

We consider the following three models of  
$N$ non-intersecting Brownian walkers on a one-dimensional line segment $[0,L]$ with different boundary conditions. Let the 
walkers be labelled by their positions at time 
$\tau$, i.e., by  
$x_1(\tau)<\ldots<x_N(\tau)$.\\

\noindent {\bf Model I:} In the first model we consider periodic boundary 
conditions on the line segment $[0,L]$. Alternatively, one can
think of the domain as a 
circle of circumference $L$ (of radius
$L/2 \pi$). All walkers start initially in the vicinity of a point on the 
circle which we call the origin. We can label the positions
of the walkers by their angles $\{\theta_1,\theta_2,\ldots, \theta_N\}$ (see 
Appendix A for details). 
Let the initial 
angles be denoted by
$\{\epsilon_1,\epsilon_2,\ldots, \epsilon_N\}$  where $\epsilon_i$'s are 
small. Eventually we will take the limit $\epsilon_i\to 0$.  
We denote by 
${R}_L^I(1)$ the reunion probability after time $\tau = 1$ (note
that the walkers, in a bunch, may wind the circle multiple times), i.e,
the probability that the walkers return to their initial positions
after time $\tau=1$ (staying non-intersecting over the time interval $\tau\in 
[0,1]$). The superscript $I$ corresponds to model I.
Evidently ${R}_L^I(1)$ depends on $N$ and also on the starting
angles $\{\epsilon_1,\epsilon_2,\ldots, \epsilon_N\}$. To avoid
this additional dependence on the $\epsilon_i$'s, let us introduce
the normalized reunion probability defined as the ratio
\begin{equation}
\label{28}
\tilde{G}_N(L)= {{R}_L^I(1) \over {R}_\infty^I(1)} \;,
\end{equation}
where we assume that we have taken the $\epsilon_i\to 0$ limit. The $\epsilon$
dependence actually cancels out between the numerator and the denominator (see
Appendix A) and hence $\tilde{G}_N(L)$ depends only on $N$ and $L$.
In Appendix A, we calculate $\tilde{G}_N(L)$ explicitly and show that
\begin{equation}\label{29}
\tilde{G}_N(L) = \frac{A_N}{L^{N^2}}\sum_{n_1=-\infty}^\infty 
\ldots \sum_{n_N = -\infty}^{\infty} 
\Delta^2(n_1,\ldots,n_N) e^{-(2 \pi^2/ L^2) \sum_{j=1}^N n_j^2} \;,
\end{equation} 
where
\begin{equation*}
\Delta(n_1,\ldots, n_N)=\prod_{1\leq i<j\leq N}(n_i-n_j) \;,
\label{vdm}
\end{equation*}
and the prefactor
\begin{equation}
A_N=\frac{1}{(2\pi)^{N/2-N^2}\prod_{j=0}^{N-1}\Gamma(j+2)}
\label{norm-Ap}
\end{equation}
ensures that $\tilde{G}_N(L\to \infty)=1$. In the next section we will
see that this normalized reunion probability $\tilde{G}_N(L)$ is,
up to a prefactor, exactly identical to the
partition function of the $2$-d Yang-Mills theory on a sphere with 
gauge group ${\rm U}(N)$. 

We remark that for the non-intersecting Brownian motions on a circle,
a similar mapping was first noticed by Minahan and Polychronakos~\cite{MP94},
with a slightly different normalization than ours. However,
the behavior of the normalized reunion probability ${\tilde G}_N(L)$ as a 
function
of the system size $L$ was not analysed in ~\cite{MP94} and consequently
they did not uncover the existence of the Tracy-Widom distribution ${\cal 
F}_2(t)$ near
the critical point $L_c(N)=2\sqrt{N}$ for large $N$ in the
reunion probability, which is indeed one of our main 
findings in this paper. \\

\noindent {\bf Model II:} In the second model the domain is the 
line segment $[0,L]$ with {\em absorbing boundary conditions at both
boundaries $0$ and $L$}. Once again, the $N$ non-intersecting Brownian
motions start initially at the positions, say, $\{\epsilon_1, 
\epsilon_2,\ldots, \epsilon_N\}$ in the vicinity of the origin where
eventually we will take the limit $\epsilon_i\to 0$ for all $i$. 
As in Model I, we define the reunion probability 
$R_L^{II}(1)$ as
the probability that the walkers return to their initial positions
after a fixed time $\tau=1$ (staying non-intersecting over the time interval
$\tau\in [0,1]$). Analogous to Model I, 
we define the normalized reunion probability
\begin{equation}\label{RL}
\tilde{F}_N(L)={R_L^{II}(1) \over R_\infty^{II}(1)} \;,
\end{equation}
which becomes independent of the starting positions $\epsilon_i$'s
in the limit when $\epsilon_i\to 0$ for all $i$.
Hence, $\tilde{F}_N(L)$ depends only on $N$ and $L$.  

\begin{figure}[ht]
\begin{center}
\includegraphics[width=10cm]{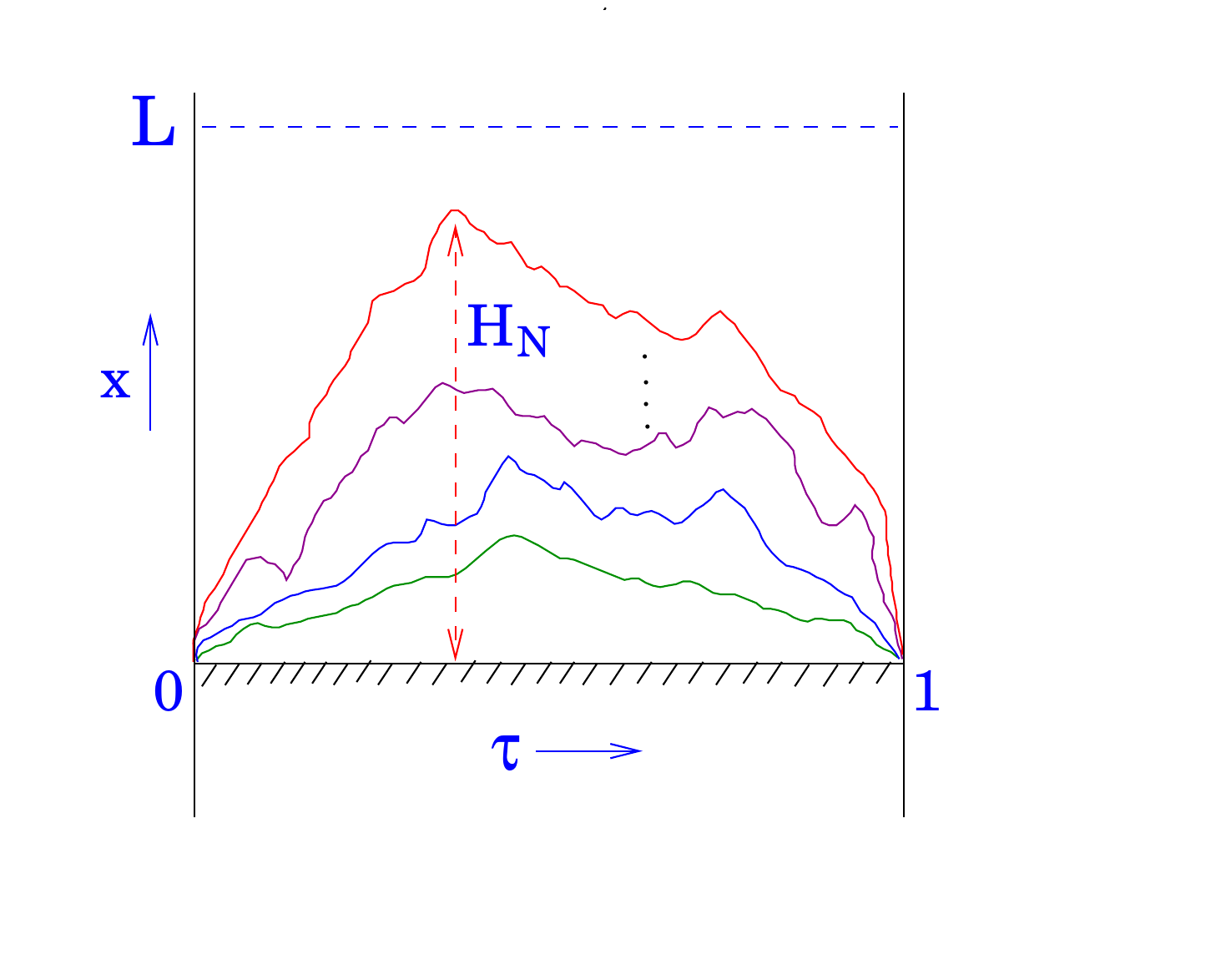}
\caption{Trajectories of $N$ non-intersecting Brownian motions 
$x_1(\tau)<x_2(\tau)<\ldots < x_N\tau)$, all start at the origin
and return to the origin at $\tau=1$, staying positive in between.
$\tilde{F}_N(M)$ denotes the probability
that the maximal height $H_N=\max_\tau\{x_N(\tau), 0\le \tau\le 1\}$ stays below 
the level $L$ 
over the time interval $0\le \tau\le 1$.}
\label{fig:watermelon}
\end{center}
\end{figure}

This ratio $\tilde{F}_N(L)$ in Model II also has a different 
probabilistic interpretation. Consider the same model but now on
the semi-infinite line $[0,\infty]$ with still absorbing boundary condition 
at $0$. The walkers, as usual, start in the vicinity of the origin 
and are conditioned to return to the origin exactly at $\tau=1$ (see Fig. 
(\ref{fig:watermelon})). 
If one plots the space-time trajectories of the walkers, a typical
configuration looks like half of a watermelon
(see Fig. (\ref{fig:watermelon})), or a watermelon in presence
of an absorbing wall. Such configurations of Brownian motions
are known as non-intersecting Brownian excursions and their
statistical properties have been studied quite extensively in
the recent past. A particular observable that has  
generated some recent interests is the so-called `height'
of the watermelon~\cite{BM2003,Fe08,Fulmek2007,KIK2008,SMCR08,KIK08} defined
as follows (see also Ref. \cite{BFPSW09} for a related quantity in the context of Dyson's Brownian motion). Let $H_N$ denote the 
maximal displacement of the rightmost walker $x_N$ in this time interval 
$\tau\in [0,1]$, i.e., the maximal height of the topmost path
in half-watermelon configuration (see Fig. (\ref{fig:watermelon})), i.e.,
$H_N=\max_\tau\{x_N(\tau),
0<\tau<1\}$. 
This global maximal height $H_N$   
is 
a random variable which fluctuates from one configuration of half-watermelon to 
another. What is the probability distribution of $H_N$? For $N=1$ the 
distribution of $H_N$ is easy to compute and already for $N=2$ it is somewhat
complicated~\cite{KIK2008}. Recently, however, an exact formula for the 
distribution of $H_N$, valid for all $N$, was derived in ~\cite{SMCR08}
using Fermionic path integral method. 

To relate the distribution of $H_N$ in the semi-infinite system
defined above to the ratio of
reunion probabilities in the finite segment $[0,L]$ defined in
\eqref{RL}, it is  
useful to consider the cumulative probability ${\rm Pr}(H_N\leq L)$ 
in the semi-infinite geometry, where
$L$ now is just a variable. To compute this cumulative probability, we need
to calculate the fraction of half-watermelon configurations (out of all
possible half-watermelon configurations) that 
never
cross the level $L$, i.e., whose heights stay below $L$ over the time interval 
$\tau\in [0,1]$ (see Fig. (\ref{fig:watermelon})). This fraction can be 
computed by putting an absorbing boundary at $L$ (thus killing all 
configurations that touch/cross the level $L$). It is then clear
that ${\rm Pr}(H_N\leq L)$ is nothing but the normalized reunion probability
${\tilde F}_N(L)$ defined in \eqref{RL}. As mentioned above, this 
cumulative probability 
distribution of the maximal height was computed exactly in Ref. \cite{SMCR08}
(see also \cite{Fe08}, \cite{KIK08} for related computations)
\begin{align}\label{Fp}
\tilde{F}_N(L)&:= {\rm Pr}(H_N\leq L) \nonumber \\
&=\frac{B_N}{L^{2N^2+N}}\sum_{n_1=1}^\infty \ldots \sum_{n_N=1}^{\infty} 
\Delta^2(n_1^2,\ldots,n_N^2)\Big(\prod_{j=1}^N n_j^2\Big)
e^{-\frac{\pi^2}{2L^2}\sum_{j=1}^N 
n_j^2} \;,
\end{align}
where
$B_N=\pi^{2N^2+N}/(2^{N^2-N/2}\prod_{j=0}^{N-1}\Gamma(2+j)\Gamma(3/2+j))$.
A brief derivation of this result is given in Appendix B.
Note that a similar probabilistic interpretation (i.e. as a cumulative 
distribution of a random variable) does not exist for the normalized reunion 
probability ${\tilde G}_N(L)$ in Model I. This is evident from the fact 
that, unlike ${\tilde F}_N(L)$ in Model II, the quantity ${\tilde G}_N(L)$ 
in Model I is not bounded from above by $1$ (this can even be seen by a direct 
computation in the $N=1$ case, see \eqref{direct_N1} in Appendix A).

In the next section, we will see that the normalized reunion probability
(or equivalently the maximal height distribution) ${\tilde F}_N(L)$ in 
Model II is, again up to a prefactor, precisely equal
to the partition function of the $2$-d Yang-Mills theory on a sphere, but 
with a gauge group ${\rm Sp}(2N)$ (as opposed to the group ${\rm U}(N)$ in 
Model I).\\

\noindent {\bf Model III:} We consider a third model of non-intersecting
Brownian motions where the walkers move again on a finite line segment 
$[0,L]$, 
but this time with {\em reflecting} boundary conditions 
at both boundaries $0$ and $L$. Again the walkers start in the 
vicinity of the origin at time 
$\tau=0$ and we consider the reunion probability $R_L^{III}(1)$ that
they reunite at time $\tau=1$ at the origin. Following Models I and II, 
we define the normalized reunion probability 
\begin{equation}\label{RLIII}
\tilde{E}_N(L)={R_L^{III}(1) \over R_\infty^{III}(1)} \;,
\end{equation} 
that is independent of the starting positions $\{\epsilon_1, 
\epsilon_2,\ldots, \epsilon_N\}$ in the limit when all the $\epsilon_i$'s
tend to zero and hence depends only on $N$ and $L$.
Following similar steps as in Models I and II, but with reflecting
boundary conditions at both $0$ and $L$, we find the  
exact expression (see Appendix B)
\begin{equation}\label{ENL}
\tilde{E}_N(L)= \frac{C_N}{L^{2N^2-N}}\sum_{n_1=-\infty}^\infty
\ldots \sum_{n_N = -\infty}^{\infty}
\Delta^2(n_1^2,\ldots,n_N^2) e^{-( \pi^2/ L^2) \sum_{j=1}^N n_j^2} \;,
\end{equation}
and the prefactor
\begin{equation}
C_N=\frac{{\pi}^{2N^2-N}\,2^{3N/2-N^2}}{\prod_{j=0}^{N-1}\Gamma(2+j)\Gamma(1/2+j)}
\label{norm-cn}
\end{equation}
ensures that $\tilde{E}_N(L\to \infty)=1$. In the next section we will
see that $\tilde{E}_N(L)$,
up to a prefactor, is exactly identical to the
partition function of the $2$-d Yang-Mills theory on a sphere with
gauge group ${\rm SO}(2N)$.

Thus, we find that by changing the boundary conditions at the edges of the
line segment $[0,L]$
in the non-intersecting Brownian motion models we can relate the normalized
reunion probability 
to the partition function of the $2$-d Yang-Mills theory on a
sphere with an appropriate gauge group. Model I, II and III 
correspond
respectively to gauge groups ${\rm U}(N)$, ${\rm Sp}(2N)$ and ${\rm SO}(2N)$.\\

\noindent {\bf Summary of new results:} Let us then summarize the main new 
results in this paper:\\

$\bullet$ We have shown that the normalized
reunion probability of a set of $N$ non-intersecting Brownian motions moving
on a line segment $[0,L]$ with a prescribed boundary condition is
identical, up to a prefactor, to the
partition function of the $2$-d Yang-Mills theory on a sphere with
an appropriate gauge group which depends on the boundary conditions
in the Brownian motion model. We have shown that three different boundary 
conditions lead respectively to the gauge groups: (I) periodic $\to$ ${\rm 
U}(N)$
(II) absorbing $\to$ ${\rm Sp}(2N)$ and (III) reflecting $\to$ ${\rm SO}(2N)$. 
\\

$\bullet$ The partition function of the $2$-d Yang-Mills theory with
a given gauge group is exactly solvable and many results are known.
In particular, it is known that in the so-called `double scaling' limit,
the singular part of the Yang-Mills free energy satisfies a Painlev\'e
equation. To use these results for the Brownian motion model via the 
correspondence established above, we need to however treat the prefactor
of the correspondence properly in the double scaling limit. Taking
into account new terms arising via these prefactors, we show that
the normalized reunion probabilities, appropriately centered and scaled 
as a function $L$ for large but fixed $N$, also share a `double scaling' 
limit where
they are precisely described by the Tracy-Widom distribution.
In the periodic case, one gets ${\cal F}_2(t)$ while for
the absorbing and reflecting cases one gets ${\cal F}_1(t)$. 
Thus, this relates for the first time (to our knowledge) the Painlev\'e
equation that appears in the Yang-Mills free energy to the Painlev\'e
equation in the Tracy-Widom distribution.\\

$\bullet$ As a byproduct of this mapping, we also show that the 
$3$-rd order phase transition between the strong and the weak coupling phases
(separated by the double scaling regime) in the $2$-d Yang-Mills
theory translates into a $3$-rd order phase transition in the behavior
of the normalized reunion probability in the Brownian motion model,
as one varies the system size $L$ across a critical value $L_c(N)\sim 
\sqrt{N}$ for large but fixed $N$.
The strong and weak coupling phases correspond respectively to the
left and right large deviation (away from the critical value $L_c(N)$) 
behaviors of the normalized reunion 
probability as a function of $L$. Again using results from the Yang-Mills 
theory (taking
into account the prefactors correctly), we thus obtain precise 
large deviation behaviors of these reunion probabilities. In particular,
in the right large deviation behavior, we find a new type of crossover 
phenomenon. \\

$\bullet$ For the special case of absorbing boundary condition (Model II),
this gives us a direct proof that the distribution of the maximal height $H_N$
of a set of $N$ non-intersecting Brownian excursions, when properly centered
and scaled, has the Tracy-Widom distribution $\mathcal{F}_1(t)$ described
in \eqref{F2-GOE}. This was shown before rather indirectly in Ref. \cite{Jo03}
via a mapping to a polynuclear growth model. Here we obtain a direct proof of 
this 
result. \\

Let us remark that while some aspects of the analogies between non-intersecting 
Brownian paths and Yang-Mills theory on the sphere have been 
noticed in earlier publications 
\cite{HT04,H05}, this precise correspondence between the normalized
reunion probability in the Brownian motion models (with different boundary 
conditions) and the partition
function of the $2$-d Yang-Mills theory on a sphere (with different gauge 
groups) seems to be new, to the best of our knowledge (except for
the periodic case when a similar correspondence was noted in ~\cite{MP94}).
More importantly, the 
probabilistic connection between the 
Yang-Mills
partition function in the double scaling limit and the 
Tracy-Widom distribution of the largest eigenvalue of a random matrix 
(established here using 
the connection via non-intersecting Brownian motions), to our 
knowledge, was not noticed earlier.

The rest of the paper is organised as follows. In Section 2, we briefly 
recapitulate the exact solution of the continuum Yang-Mills theory 
in two dimensions and then establish the correspondence between
the partition functions of the gauge theory with normalized
reunion probabilities in the Brownian motion models defined
above. Next we study the consequences of this correspondence
for Model I and Model II respectively in Sections 3 and 4.
In particular, we will see how the Tracy-Widom distributions
${\cal F}_2(t)$ and ${\cal F}_1(t)$ emerge in the double scaling limit 
in Models I and II respectively. For Model 
II, this correspondence
also provides us with the precise asymptotic distribution of the
maximal height for $N$ non-intersecting Brownian excursions. We do not discuss 
Model III in details here as the analysis is very similar to that of Model 
II. 
The detailed derivation of the expression of ${\tilde G}_N(L)$ in 
\eqref{29} is provided in the Appendix A. The derivations of ${\tilde F}_N(L)$
in \eqref{Fp} and of ${\tilde E}_N(L)$ in \eqref{ENL} follow via
similar calculations which are briefly outlined in Appendix B. 

\section{Correspondence between $2$-d Yang-Mills theory and non-intersecting
Brownian motions on a line segment $[0,L]$}

We start by briefly recapitulating how one computes the $2$-d Yang-Mills
partition function~\cite{M75,Ru90,G93} (see also the review by
Cordes, Moore and Ramgoolam~\cite{CMR95}). Consider a 
general orientable
two-dimensional manifold $\mathcal{M}$ with corresponding volume form
$\sqrt{g}$. At each point $x$ of this manifold sits a
gauge field $A_\mu(x)$ (with the space index $\mu=1,\, 2$) which
is an $(N\times N)$ matrix. For simplicity, let us first consider
pure ${\rm U}(N)$ Yang-Mills gauge theory whose  
partition function in continuum is defined by the
functional integral
\begin{equation}
\label{YM2-pf}
\mathcal{Z}_ \mathcal{M} =\int[\mathcal{D}A_\mu] 
e^{-(1/4\lambda^2)\int_ \mathcal{M}\sqrt{g}\, {\rm Tr}[F^{\mu \nu}
F_{\mu \nu}] d^2x} \;,
\end{equation}
where $\lambda$ is a coupling constant and the field strengths are defined by 
\begin{equation}
F_{\mu \nu}= \partial_\mu A_{\nu}-\partial_{\nu}A_{\mu} +i[A_\mu,A_\nu] \;.
\label{field-YM}
\end{equation}
Under a local gauge transformation $A_\mu\to S^{-1}(x) A_\mu S(x)-i 
S^{-1}(x) 
\partial_\mu S(x)$ (where $S(x)$ is an $(N\times N)$ unitary matrix), the
field strengths transform as $F_{\mu \nu}\to S^{-1}(x)F_{\mu \nu} S(x)$
thus keeping the action in \eqref{YM2-pf} gauge invariant.

\begin{figure}[ht]
\begin{center}
\includegraphics[width=10cm]{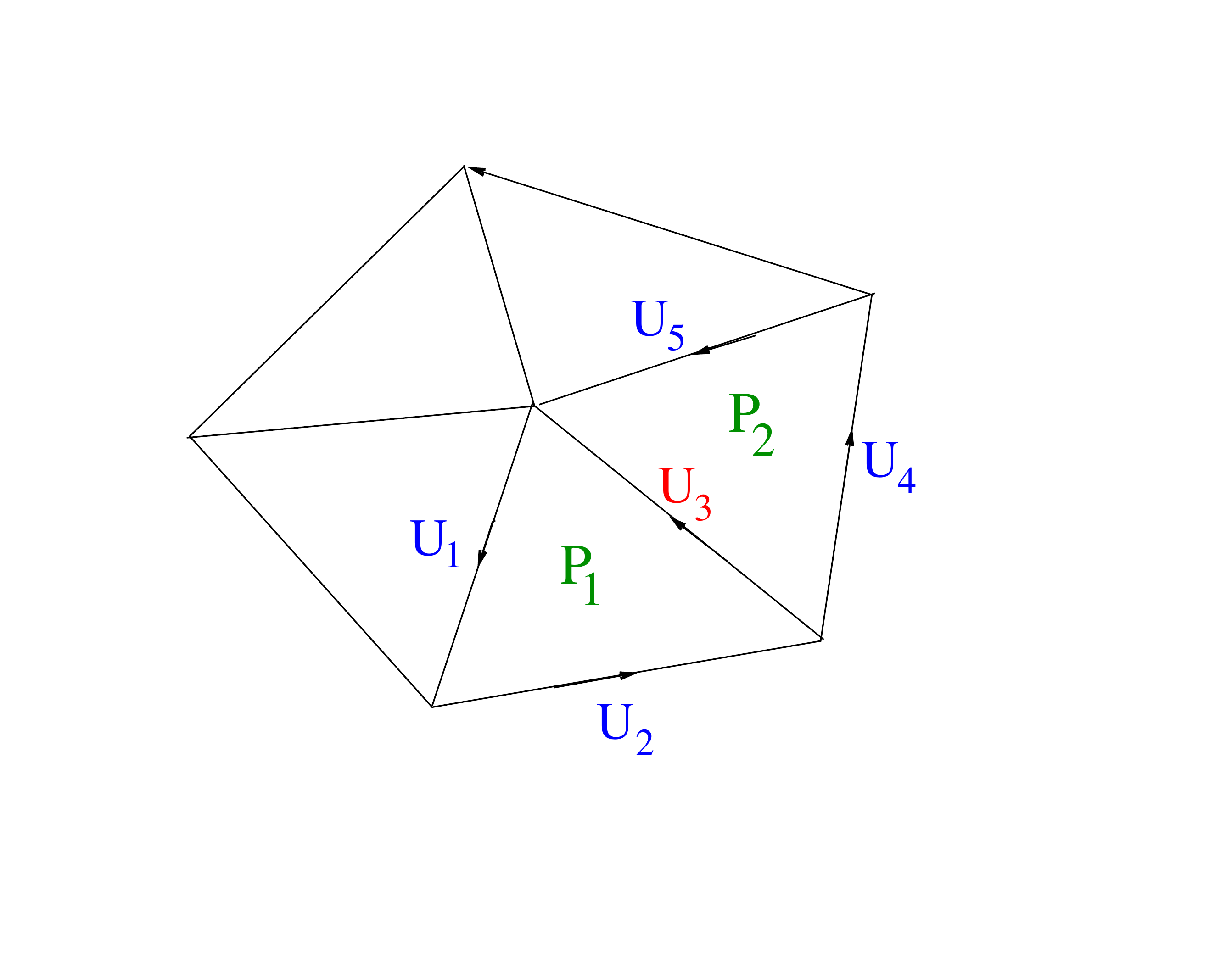}
\caption{a typical triangulation of the two-dimensional manifold with
unitary matrices $U_L$'s on the edges. One can fuse two triangles $P_1$
and $P_2$ by integrating the matrix $U_3$ along their common edge and
obtain a parallelogram with four edges with matrices $U_1$, $U_2$, $U_4$
and $U_5$ on these edges.} 
\label{fig:triangle}
\end{center}
\end{figure}

The partition function in \eqref{YM2-pf} in two dimensions can be computed 
exactly via the original idea due to Migdal~\cite{M75}. One can actually use
a particular lattice regularization of the continuum theory which is both exact 
and additive in the following sense~\cite{G93}. One can divide the 
manifold into polygons (for example one can choose triangles as basic 
units or plaquettes) and define a unitary matrix $U_L$ sitting at the
center of each link $L$ of this triangulated manifold (see Fig. 
\ref{fig:triangle}). Then 
the lattice regularized partition function can be written~\cite{G93}
\begin{equation}
\mathcal{Z}_ \mathcal{M}({\rm lattice})= \int \prod_{L} dU_L \prod_{\rm 
plaquettes} Z_P[U_P],
\label{lattice-pf}
\end{equation}
where $U_P= \prod_{L\in {\rm plaquette}} U_L $ is called the loop product
introduced by Wilson~\cite{Wilson74}
and $Z_P(U_P)$ 
is some appropriate
lattice action associated with the plaquette $P$. The only constraint
on the choice of $Z_P$ is that it should reduce to the continuum action
when the plaquette size goes to zero. 

A standard choice for $Z_P(U_P)$
is the Wilson action~\cite{Wilson74}
\begin{equation}
Z_P(U_P)= \exp\left[b\, N {\rm 
Tr}(U_P+U_P^{\dagger})\right] \;,
\label{Wilson-action}
\end{equation}
which does reduce to the continuum action in the limit when the plaquette
area goes to zero.
With this choice, the lattice partition 
function in
\eqref{lattice-pf} is exactly solvable~\cite{GW80,Wadia80} as it reduces
to computing a single matrix integral in Eq. (\ref{firsteqn}) already
discussed in the introduction. However, the Wilson action is not invariant
under renormalization in the following sense. Following Migdal~\cite{M75},  
one can take two 
adjacent 
plaquettes $P_1$ and $P_2$ with their respective actions $Z_{P_1}$
and $Z_{P2}$ and fuses them to form a bigger plaquette with area
equal to that of $P_1+P_2$, after
integrating out the unitary matrix sitting on the common link
between the two plaquettes (see Fig. \ref{fig:triangle}). This
gives the Migdal recursion relation
\begin{equation}
\int dU_3 Z_{P_1}(U_1U_2U_3)Z_{P_2}(U_4U_5U_3^{\dagger})= Z^{\prime}_{P_1+P_2}  
(U_1U_2U_4U_5) \;.
\label{recur:migdal}
\end{equation}

In general, the renormalized plaquette action $Z^{\prime}_P$ does not have
the same functional form as the bare action $Z_P$ (which is
indeed the case when one chooses
Wilson action as the bare action). So, the natural thing to look for is
the fixed point solution
of this recursion relation that keeps the form of $Z_P$ invariant
under renormalization. Indeed, Migdal~\cite{M75} and later 
Rusakov~\cite{Ru90} showed that the appropriate fixed point action is given by 
the so-called heat-kernel action
\begin{equation}
Z_P= \sum_{R} d_R \chi_R(U_P) \exp\left[-\frac{ {\tilde 
{\lambda}}A_P}{2N}\, C_2(R)\right] \;,
\label{heat-kernel_1}
\end{equation}  
where the sum runs over all irreducible representations $R$ of ${\rm U}(N)$, $d_R$
is the dimension of $R$, $\chi_R(U_P)$ is the character of $U_P$ in the 
representation $R$, $C_2(R)$ is the quadratic Casimir operator of $R$ and $A_P$
is the area of the plaquette. The coupling constant ${\tilde
{\lambda}}$ is fixed and will henceforth be chosen to be unity:
${\tilde {\lambda}}=1$.
One can 
also verify 
that this heat-kernel
action reduces to the continuum action in the limit when the area of 
each basic plaquette goes to zero~\cite{M75,Ru90}. Note that this fixed 
point choice of $Z_P$ makes the lattice representation of the continuum 
theory exact in the sense that the final result is independent of 
the triangulation as one can add as many triangles (and in whichever way)
to cover the manifold thus approaching the continuum limit~\cite{G93}.
The name heat-kernel stems from the fact that one can express 
the heat-kernel action $Z_P(U)=\langle 
U|\exp[-(1/2N)A 
\Delta]|1\rangle$ where $\Delta$ is the Laplacian on the group~\cite{MO81}. 
This fact 
already gives a hint that there might be an underlying diffusion process 
inbuilt in this effective action, though the precise fact that they correspond
to non-intersecting Brownian motions is still not evident at this point.
Using the heat-kernel action $Z_P$ in \eqref{heat-kernel_1} one can then 
evaluate the full partition function on the manifold~\cite{Ru90}
\begin{equation}  
\mathcal{Z}_ \mathcal{M}= \sum_R 
d_R^{2-2p}\exp\left[-\frac{A}{2N}\,C_2(R)\right] \;,
\label{htpf}
\end{equation}
where $p$ is the genus of the manifold and $A$ is the total surface area.
This then represents an exact solution of the partition function 
of the continuum Yang-Mills theory on a twodimensional manifold.
Note that even though we have specifically used the group ${\rm U}(N)$ for the 
discussion above, the result \eqref{htpf} is valid as well for other groups
such as ${\rm SU}(N)$, ${\rm Sp}(N)$ and ${\rm SO}(N)$. 

For a sphere, using $p=0$, one gets 
\begin{equation}
\mathcal{Z}_ \mathcal{M}= \sum_R 
d_R^{2}\exp\left[-\frac{A}{2N}\, C_2(R)\right].
\label{pfsphere}
\end{equation}

The irreducible representations can be labelled by the lengths of the Young 
diagrams and one can explicitly express the partition function as a
multiple sum. For the group ${\rm U}(N)$, the partition function 
reads~\cite{G93,GM94}
\begin{equation} 
\label{ZU}
\mathcal{Z}_{\mathcal{M}}:=\mathcal{Z}(A;{\rm U}(N))= 
c_N e^{-A\frac{N^2-1}{24}}\sum_{n_1,\ldots,n_N=-\infty}^\infty 
\Delta^2(n_1,\ldots,n_N)e^{-(A/2N)\sum_{j=1}^N n_j^2} \;,
\end{equation}
where $\Delta(n_1,n_2,\ldots, n_N)$ is the van der Monde determinant defined
in \eqref{vdm} and $c_N$ is a constant independent of $A$.
For the groups ${\rm Sp}(2N)$ and ${\rm SO}(2N)$ one can similarly express the
partition function as a multiple sum~\cite{CNS96}. For example,
for ${\rm Sp}(2N)$ one gets
\begin{equation}
\label{ZSpN}
 \mathcal Z(A;{\rm Sp}(2N)) = \hat{c}_N e^{A\,(N+\frac{1}{2})\frac{N+1}{12}}
 \sum_{n_1,\dots,n_N = - \infty}^\infty
 \Delta^2(n_1^2,\dots,n_N^2) 
\Big ( \prod_{j=1}^N n_j^2 \Big ) e^{- {A \over 4 N} \sum_{j=1}^N n_j^2},
 \end{equation}
where $\hat{c}_N$ is independent of $A$. 
Similarly, for the ${\rm SO}(2N)$ group, one obtains~\cite{CNS96}
\begin{equation}
\label{ZSON}
\mathcal Z(A;{\rm SO}(2N))= \hat{b}_N e^{A\, (N-\frac{1}{2})\frac{N-1}{12}}
\sum_{n_1,\dots,n_N = - \infty}^\infty
 \Delta^2(n_1^2,\dots,n_N^2)
e^{- {A \over 4 N} \sum_{j=1}^N n_j^2},
 \end{equation}
where $\hat{b}_N$ is independent of $A$.

Comparing the formulas 
in \eqref{29}, \eqref{Fp} and \eqref{ENL} with the expression for 
partition functions respectively in
\eqref{ZU}, \eqref{ZSpN}, and \eqref{ZSON}, we see that the
normalized reunion probabilities 
in the three models of non-intersecting Brownian motions, up to 
prefactors that can be computed explicitly, are
isomorphic to the partition functions of the Yang-Mills theory on
a sphere with respective gauge groups ${\rm U}(N)$, ${\rm Sp}(2N)$ and ${\rm 
SO}(2N)$,
provided we make the identification $A/2N \to 2\pi^2/L^2$ in Model I,
$A/4N\to \pi^2/{2L^2}$ in model II and $A/4N\to \pi^2/L^2$ in model III.
More precisely, up to known prefactors,
\begin{eqnarray}
\tilde{G}_N(L) &\propto & \mathcal Z\left(A=\frac{4\pi^2N}{L^2};{\rm 
U}(N)\right) \;, \label{corrUN} \\
\tilde{F}_N(L)  &\propto & \mathcal Z\left(A=\frac{2\pi^2N}{L^2};{\rm
Sp}(2N)\right) \;, \label{corrSP2N} \\
\tilde{E}_N(L)&\propto & \mathcal Z\left(A=\frac{4\pi^2N}{L^2};{\rm
SO}(2N)\right) \;. \label{corrSO2N} 
\end{eqnarray}
This correspondence between the normalized reunion probability 
for non-intersecting Brownian motions
with different boundary conditions and the Yang-Mills partition functions 
on a sphere with corresponding gauge groups is 
one 
of the main 
observations of this paper.
  
In the next two sections we study the consequences of this correspondence
in detail for Model I and II. We skip detailed studies
of Model III since it can be handled exactly in the same way as 
Model II. The main point is to derive the precise asymptotics of the 
normalized reunion probabilities in the two models (in particular for Model 
II this will give us the asymptotic behavior of the distribution of the maximal
height $H_N$ for non-intersecting Brownian excursions) by using the known
behavior of the asymptotic properties of the partition functions of the 
corresponding gauge theory, albeit taking into account correctly the $L$ and $N$ 
dependence
of the respective prefactors.  
 
\section{Brownian walkers on a circle: Model I}

Comparing \eqref{ZU} and \eqref{29} we have the following exact identity 
between
the normalized reunion probability and the ${\rm U}(N)$ Yang-Mills partition 
function on a sphere (upon substituting $A=4\pi^2N/L^2$ in \eqref{ZU})
\begin{equation}
\tilde G_N(L)= \frac{A_N}{c_N}\, e^{\pi^2 N(N^2-1)/{6L^2}}\, L^{-N^2}\, 
\mathcal{Z}\left( \frac{4\pi^2 N}{L^2};{\rm U}(N)\right) \;,
\label{G_Z}
\end{equation}
where the constants $A_N$, given in \eqref{norm-Ap}, is
independent of $L$. Similarly the constant $c_N$ is independent
of $L$ and can be fixed in any way. Later, we choose
$c_N$ such that
for large $N$, $\log c_N \simeq -N^2 \log N$. This choice of $c_N$
ensures that the free energy $\log \mathcal{Z}_{\mathcal{M}}\sim {\cal O}(N^2)$
for large $N$ (see below).

\begin{figure}
\begin{center}
 \includegraphics[width = \linewidth]{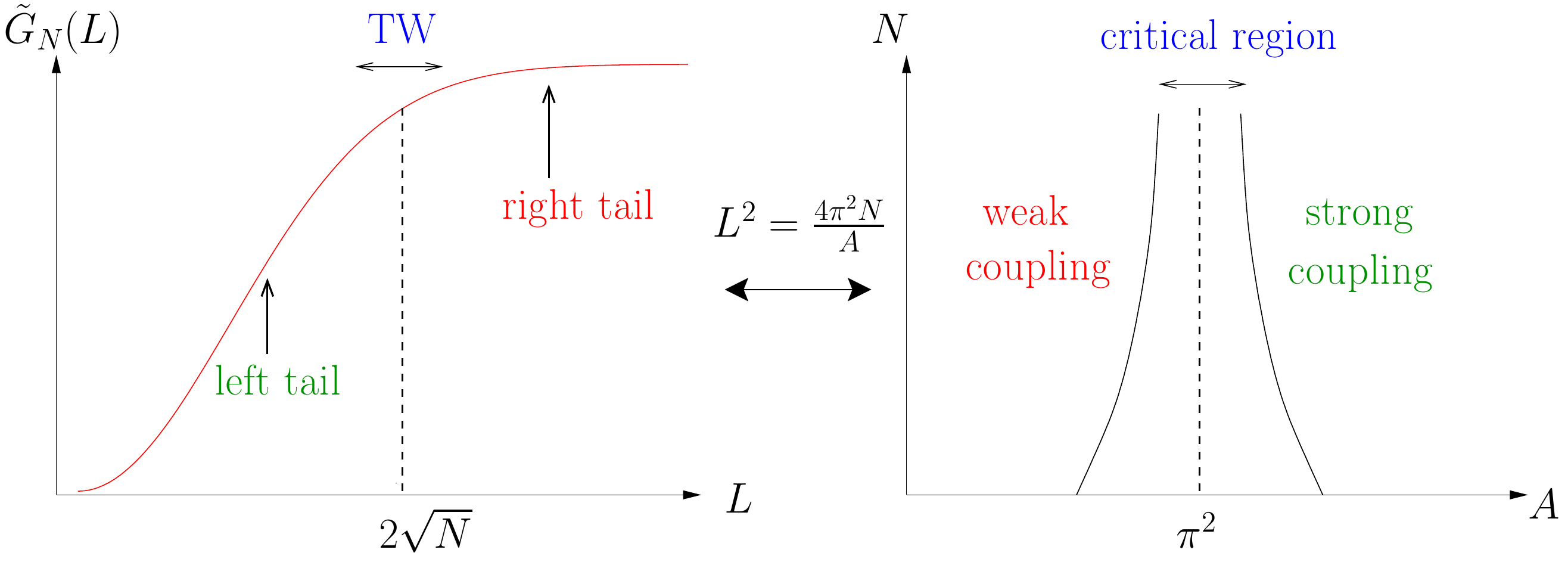}
\caption{{\bf Left:} 
Schematic sketch of $\tilde G_N(L)$ as defined in Eq. (\ref{28}) for $N$ 
vicious 
walkers on a circle, as a function of $L$, for fixed but large $N$.  {\bf 
Right:} Sketch of the phase 
diagram in the plane 
$(A,N)$ of two-dimensional Yang-Mills theory on a sphere with the gauge 
group ${\rm U}(N)$ as obtained in Ref. \cite{DK93}. The weak (strong)
coupling phase in the right panel corresponds to the right (left)
tail of $\tilde G_N(L)$ in the left panel. The critical region around
$A=\pi^2$ in the right panel corresponds to the Tracy-Widom (TW) regime
in the left panel around the critical point $L_c(N)= 2\sqrt{N}$.}   
\label{fig_corres}
\end{center}
\end{figure}

In Ref. \cite{DK93} it was shown that for large $N$, $\mathcal{Z}(A;{\rm 
U}(N))$ exhibits
a $3$-rd order phase transition at the critical value $A = \pi^2$ 
separating a weak coupling 
regime for $A < \pi^2$ and a strong coupling 
regime for $A > \pi^2$ (see Fig. \ref{fig_corres}). This is the so
called Douglas-Kazakov phase transition~\cite{DK93}. 
Using the correspondence 
$L^2 := 4 \pi^2 N/A$, we then find that ${\tilde G}_N(L)$, considered
as a function of $L$ with $N$ large but fixed, must also exhibit
a $3$-rd order phase transition at the critical value $L_c(N)=2 \sqrt{N}$. 
Furthermore, the weak coupling regime ($A<\pi^2$) 
corresponds to $L > 2 \sqrt{N}$ and thus 
describes the right tail of $\tilde G_N(L)$, 
while the strong coupling regime corresponds to 
$L < 2 \sqrt{N}$ and describes instead the left tail 
of $\tilde G_N(L)$ (see Fig. \ref{fig_corres}). 
The critical regime around $A = \pi^2$ is the so-called 
"double scaling limit" in the matrix model and has width of 
order $N^{-2/3}$. It corresponds, as we will see, to the region 
(of width $N^{-1/6}$ around $L=2 \sqrt{N}$) where $\tilde G_N(L)$, 
correctly shifted and scaled, is described by the 
Tracy-Widom distribution ${\cal F}_2(t)$.    

\subsection{Large deviation tails}

Let us first consider the behavior of $\tilde G_N(L)$, as a function of $L$,
in the two tails far away from the critical value $L_c(N)=2 \sqrt{N}$ for
large but fixed $N$.   
To do this, we can exploit the exact identity
in \eqref{G_Z} and use the known properties of the ${\rm U}(N)$ partition
function respectively in the strong ($A> \pi^2$) and the weak ($A< 
\pi^2$) coupling phases. The left ($L< 
2\sqrt{N}$) and the right ($L> 2\sqrt{N}$) tails of $\tilde G_N(L)$,
away from the critical point $L_c(N)=2\sqrt{N}$, 
correspond respectively to the behavior of the partition
function in the strong ($A>\pi^2$) and weak ($A<\pi^2$) coupling phases.

Let us begin by summarizing the known properties of the ${\rm U}(N)$
partition function in the strong and weak coupling phases.
The summand in \eqref{ZU} can naturally be regarded as a function 
of 
$n_i/N$ ($i=1,\dots,N$).
In the large $N$ limit the variables $\{n_i/N\}_{i=1,\dots,N}$ approximate the 
coordinates of a continuous $N$ particle system. Associated with the particles 
is a density $\rho(x)$, but because the
lattice spacing is $1/N$ and there are  $N$ particles 
the density at any point cannot exceed 1, and thus
$\rho(x) \le 1$ for all $x$. By using this viewpoint to 
perform a constrained saddle point analysis of
(\ref{ZU}) (see also Section \ref{s3a} below), it was shown by Douglas and 
Kazakov \cite{DK93} that 
\begin{equation} \label{ZA1}
\lim_{N \rightarrow \infty}\frac{1}{N^2}\log \mathcal{Z}(A;{\rm U}(N))=\left\{ 
\begin{tabular}{ll}
$F_-(A)$, & $A\leq \pi^2$ \;,  \\
$F_+(A)$, & $A\geq \pi^2$ \;,
\end{tabular}
\right. 
\end{equation}
where 
\begin{equation}
F_-(A) = - {3 \over 4} - {A \over 24} - \frac{1}{2}\log A \;.
\end{equation}
The function $F_+(A)$ about $A=\pi^2$ agrees with $F_-(A)$ up to ${\cal O}((A-\pi^2)^2)$ in its power series expansion. Thus \cite[eq.~(40)]{DK93}
\begin{equation} \label{34}
F_+(A) - F_-(A) = {1 \over 3 \pi^6} (A - \pi^2)^3 + {\cal O}((A-\pi^2)^4) \;.
\end{equation}
Its explicit form --- or more precisely that of its derivative --- is given in terms of elliptic integrals.
Specifically,
with $K \equiv K(k)$, $E \equiv E(k)$ denoting standard elliptic integrals, where $k$ is specified by the requirement 
\begin{equation}
\frac{A}{4}=(2E-k'^2K)K \;, \; k'^2=1-k^2 \;,
\end{equation}
and
\begin{equation}
a=\frac{4K}{A} \;,
\end{equation}
one has  \cite[eq.~(35)]{DK93}
\begin{equation}
F_+'(A)=\frac{a^2}{6}-\frac{a^2k'^2}{12}-\frac{1}{24}+\frac{a^4k'^4A}{96} \;.
\end{equation} 

In (\ref{ZA1}), the case $A < \pi^2$ corresponds to the
weak coupling phase and the 
constraint on the
density of the variables $n_i/N$ being less than 1 can be ignored, implying 
that the Riemann
sum is well approximated by the corresponding multi-dimensional integral. 
In the latter
the $A$ dependence  can be determined by scaling, and the resulting integral 
evaluated
explicitly (see e.g.~\cite[eq.~(4.140)]{Fo10}) to deduce the stated result. 
However for
$A > \pi^2$ doing this would imply that the density is greater than 1, 
so the discrete sum is no longer
well approximated by the continuous integral. The density saturates at 1 for 
some range of values
of the variables $n_i/N$ about the origin, but goes continuously to zero to be 
supported on
a finite interval (see e.g.~Fig.~2 in \cite{AAA05}).

Let us now use these known results on the partition function
in our exact identity \eqref{G_Z} to derive the corresponding
large deviation properties of $\tilde G_N(L)$. Setting, for
convenience, $L=2\sqrt{N}r$ in \eqref{G_Z} so that $r=1$ corresponds to the 
critical point, we get 
\begin{equation}
\label{18}
\tilde{G}_N(2\sqrt{N}r)=\frac{A_N e^{\pi^2 
(N^2-1)/24r^2}}{c_N (2\sqrt{N}r)^{N^2}}\mathcal{Z}\Big(\frac{\pi^2}{r^2};{\rm U}(N) 
\Big). 
\end{equation}
We first note that $A_N$ in Eq. (\ref{norm-Ap}) contains the product
$\prod_{j=0}^{N-1}\Gamma(j+2)=G(N+2)$ in the denominator, 
where $G(x)$ denotes the 
Barnes $G$-function whose asymptotic expansion is 
(see e.g., \cite[eq.(4.184)]{Fo10})
\begin{equation}
\ln (G(z + 1)) = \frac{z^2\ln z}{2} - \frac{3z^2}{4} + 
\frac{\ln(2\pi)}{2} z - \frac{1}{12}\ln(z) + \zeta'(-1) + {\cal O}(1/z) \;.
\label{barnesg}
\end{equation}
Using this result and the choice $c_N\simeq e^{-N^2 \log N}$, we see
that the leading $N^2\log N$ term cancels when one considers the ratio 
\begin{equation}
\frac{A_N e^{\pi^2 (N^2-1)/24r^2}}{c_N (2 \sqrt{N} 
r)^{N^2}}\mathop{\sim}e^{N^2(3/4-\log (r /\pi)+ \pi^2/24 r^2)+{\cal 
O}(N)}. 
\end{equation}
Consequently, the result in
Eq. (\ref{ZA1}) gives the large deviation formula
\begin{equation} \label{19}
\tilde{G}_N(2\sqrt{N}r)\mathop{\sim}\left\{ 
\begin{tabular}{ll}
$1,$ & $r\geq1$ \;, \\
$e^{-N^2(F_-(\pi^2/r^2)-F_+(\pi^2/r^2))}$, & $r\leq1$.
\end{tabular}
\right. 
\end{equation}
This is then a new result on the far tails of the normalized reunion 
probability $\tilde G_N(L)$, as a function of $L$, for fixed but large $N$.

Note that the precise meaning of $\sim$ in \eqref{19} is the following
\begin{equation}
\label{precise-19}
-\lim_{N\to \infty} \frac{1}{N^2} \ln\left[ 
\tilde{G}_N(2\sqrt{N}r)\right]=\left\{
\begin{tabular}{ll}
$0,$ & $r\geq1$  \\
$F_-(\pi^2/r^2)-F_+(\pi^2/r^2)$, & $r\leq1$.
\end{tabular}
\right.
\end{equation}
Thus this calculation only gives the leading ${\cal O}(N^2)$ term. While in the
left tail ($r\le 1$) the leading term of ${\cal O}(N^2)$ is a finite
nontrivial quantity, this leading ${\cal O}(N^2)$ term actually vanishes
in the right tail $(r\ge 1)$. So, to obtain the finer behavior
in the right tail one needs to keep track of the next subleading correction
of ${\cal O}(N)$ term in $\ln[ \tilde{G}_N(2\sqrt{N}r)]$ for $r\ge 1$. 

In the gauge theory context, this means that we need to
obtain in the weak coupling phase, not only the leading term
of ${\cal O}(N^2)$ but also the subleading corrections. 
Fortunately, these subleading corrections in the free energy $\ln \mathcal{Z}$
in the weak coupling phase were also calculated by Gross and 
Matytsin~\cite[eq.~(2.41) ]{GM94}
\begin{eqnarray}
\label{in}
\log  \mathcal{Z}(A;{\rm U}(N)) &\sim& N^2 F_-(A) + {A \over 24} - {(-1)^N \over \sqrt{2 \pi N}}
{A \over 2 \pi^2} \Big ( 1 - {A \over \pi^2} \Big )^{-1/4} e^{- \frac{2 \pi^2 N}{A}\gamma(\frac{A}{\pi^2})} \nonumber \\
&+& {\cal O}\left(e^{-\frac{4 \pi^2 N}{A} \gamma\left(\frac{A}{\pi^2} \right)} \right) \;,
\end{eqnarray}
up to terms independent of $A$ (these will depend on the precise form of $c_N$ in 
(\ref{ZU}))
where
\begin{equation}\label{gx}
\gamma(x) = \sqrt{1 - x} - {x \over 2} \log {1 + \sqrt{1 - x} \over 1 - \sqrt{1 - x}} = {2 \over 3}
(1-x)^{3/2} + {\cal O}((1-x)^{5/2}).
\end{equation}
We can then use this result in our identity \eqref{18} to predict the
following right large deviation tail 
\begin{equation}\label{ld}
1-\tilde{G}_N(2\sqrt{N}r) \: \sim \: (-1)^N e^{-2  N r^2  \gamma(1/ r^2)}, 
\qquad r > 1,
\end{equation}
which shows an interesting oscillatory behavior. Again this is
a new result for the normalized reunion probability of 
the non-intersecting Brownian motions on a circle. Let us
recall that $\tilde G_N(L)$ does not have the meaning of a 
cumulative distribution and hence the oscillating sign of 
$[1-\tilde{G}_N(2\sqrt{N}r)]$ (with $N$) is not really problematic. 

\subsection{Double scaling limit}\label{sdd}

Having obtained the precise large $N$ asymptotic behavior of $\tilde G_N(L)$ as 
a function of $L$
in the left $(L < 2\sqrt{N})$ and the right $(L > 2\sqrt{N})$ tails, we now
turn our attention to the 
behavior of $\tilde G_N(L)$ in the vicinity of the critical point, i.e.,
when $L$ is close to $L_c(N)= 2\sqrt{N}$. In the gauge theory, this
corresponds to the double scaling regime near the critical point
$A=\pi^2$. Below, we first discuss the known results on the partition function
in the double scaling regime. We then use these results in our exact
identity \eqref{G_Z} to show that correspondingly 
$\tilde G_N(L)$, in a narrow region $|L-2\sqrt{N}|\sim N^{-1/6}$
around the critical point $L_c(N)=2\sqrt{N}$, has the
scaling behavior 
\begin{equation}
\tilde G_N(L) \to {\cal F}_2\left(2^{2/3} N^{1/6}|L-2\sqrt{N}|\right) \;,
\label{scaling1}
\end{equation}
where the scaling function ${\cal F}_2(t)$ is precisely the Tracy-Widom
distribution function for GUE random matrices defined by 
\begin{equation}
\label{TWF2t}
{\cal F}_2(t)=
\exp\bigg(-\int_t^\infty (s-t)q^2(s)ds\bigg) \;,
\end{equation}
with $q(s)$ satisfying the Painlev\'e equation \eqref{qdiff}.
This is the main result of this subsection and details are
provided below.

Gross and Matytsin \cite{GM94} used the method of orthogonal polynomials 
(see e.g.~\cite[Ch.~5]{Fo10}) to analyse the $N \to \infty$ asymptotic 
behaviour of
(\ref{ZU}) when the coupling $A$ is tuned in the neighbourhood of the 
transition point
$A = \pi^2$. Explicitly, in what is referred to as the double scaling limit,
this is achieved by taking  
$N \to \infty$ while keeping $(\pi^2 - A) N^{2/3}$ fixed.

As most clearly set out in \cite[eq.~(23)]{CNS96}, for certain 
orthogonal polynomial
normalizations $R_N^{(\pm)}$, the method of orthogonal polynomials gives
\begin{equation}\label{RN}
{d^2 \over d A^2} \log  \mathcal{Z}(A;{\rm U}(N)) = {1 \over 4 N^2} R_N^{(\pm)} (
R_{N+1}^{(\pm)} + R_{N-1}^{(\pm)})
\end{equation}
with $+$ ($-$) chosen according to $N$ being even (odd). Moreover, it is shown in
\cite{GM94}  that for $j$ near $N$
\begin{equation}\label{Rj}
R_j^{(\pm)} = {N^2 \over \pi^2} \mp (-1)^j N^{5/3} f_1(x) + {\cal O}(N^{4/3})
\end{equation}
where
\begin{equation}\label{Rx}
x = N^{2/3} \Big ( 1 - {A \over \pi^2} \Big ).
\end{equation}
In (\ref{Rj}) $f_1$ satisfies the differential equation
\begin{equation}\label{11a}
f_1'' - 4x f_1 - {\pi^4 \over 2} f_1^3 = 0 \;,
\end{equation}
subject to the boundary condition
\begin{equation}\label{12a}
f_1(x) \mathop{\sim}\limits_{x \to \infty} - \sqrt{2 \over \pi^5} {e^{-{4 \over 3} x^{3/2}} \over x^{1/4}} \;,
\end{equation}
(here we have corrected a factor of $\pi^2$ in Eq.~(5.13) of \cite{GM94} in writing (\ref{11a}),
while in writing (\ref{12a}) we have changed the sign in Eq.~(5.14); to see that the latter is needed
compare Eq.~(5.14) with Eq.~(5.4)).

As noted in  \cite{GM94}, (\ref{11a}) can be identified with the Painlev\'e II equation 
(\ref{forthequation2}) in the case $\alpha = 0$. Explicitly, the latter is obtained upon the
substitutions
\begin{equation}\label{xu}
x = 2^{-2/3} t, \qquad f_1(x) = - {2^{5/3} \over \pi^2} u(t)
\end{equation}
and furthermore, upon recalling the asymptotic expansion
$$
{\rm Ai} (t) \mathop{\sim}\limits_{t \to \infty}  {e^{-2 t^{3/2}/3} \over 2 \sqrt{\pi} t^{1/4}} \;,
$$
 a boundary condition consistent with (\ref{bc}) is obtained.

Substituting (\ref{Rj}) in (\ref{RN}) implies that with $x$ as specified by (\ref{Rx}) fixed
\cite[below (5.16)]{GM94}
$$
{d^2 \over d A^2} \log  \mathcal{Z}(A;{\rm U}(N)) \sim {N^2 \over 2 \pi^4} \Big (
1 - {2 x \over N^{2/3}} - { \pi^4 \over 2 N^{2/3}} f_1^2(x) \Big ) \;,
$$
or equivalently, upon recalling (\ref{Rx}) and making use of the variables (\ref{xu}),
\begin{equation}
{d^2 \over d t^2} \log e^{-A^2 N^2/ 4 \pi^4}  \mathcal{Z}(A;{\rm U}(N))  \Big |_{A = \pi^2 -
t \pi^2/(2N)^{2/3}} = - q^2(t) \;,
\end{equation}
where $q(t)$ is as in (\ref{fortheqn}).

The relation (\ref{18}) now tells us that for the ratio of return probabilities we have
that for $N \to \infty$
\begin{equation}\label{32}
{d^2 \over d t^2} \log \tilde{G}_N(2  \sqrt{N} (1 + t/2(2 N)^{2/3}) ) = - q^2(t).
\end{equation}
But the distribution (\ref{F2}) for the scaled largest eigenvalue in the GUE
satisfies this same relation, and so we have
\begin{equation}\label{33}
\lim_{N \to \infty}  \tilde{G}_N(2  \sqrt{N} (1 + t/2(2 N)^{2/3}) ) = {\mathcal F}_2(t).
\end{equation}
Here use has also been made of the fact, which follows from (\ref{19}), that the left-hand side (lhs) tends to
1 as $t \to \infty$.

We know from previous studies of large deviation formulas
associated with the largest eigenvalue of a random 
matrix~\cite{DM06,DM08,VMB07,MV09,NM09,BEMN10}, where the transition region is
specified by the Tracy-Widom scaling function, that the expansion of the large 
deviation functions 
around the transition
point coincides with the tail behaviours of the transition region 
Tracy-Widom scaling function.
Interestingly, this property
holds in the present setting for one side (left) of the tails only.
Thus making use of the expansions (\ref{34}) and
(\ref{gx}) we obtain
\begin{align}
& \tilde{G}_N(2 \sqrt{N} (1 + t/2(2N)^{2/3}) \mathop{\sim}\limits_{t \to - \infty}
e^{{t^3\over 12} } \\
& 
 \tilde{G}_N'(2 \sqrt{N} (1 + t/2(2N)^{2/3})  \mathop{\sim}\limits_{t \to \infty}
 e^{- { 2t^{3/2} \over 3}}
 \end{align}
 with the first of these only the precise  tail form ${\mathcal F}_2(t)$ \cite{TW94a}.
 The $t \to \infty$ tail of ${\mathcal F}_2(t)$ has the leading form
 $e^{- {4 t^{3/2} \over 3}}$, and so differs by a factor of 2 in the exponent. 
This is the indication of an interesting crossover which we describe here qualitatively \cite{moredetails}. Close to $L = 2 \sqrt{N}$, with $L - 2 \sqrt{N} \sim {\cal O}(N^{-1/6})$, a calculation of $\tilde G_N(2  \sqrt{N} (1 + t/2(2 N)^{2/3}) )$ beyond leading order shows that
\begin{eqnarray}\label{beyond}
\log \tilde{G}_N(2  \sqrt{N} (1 + t/2(2 N)^{2/3}) ) = \log{\left({\cal F}_2(t)\right)} + N^{-1/3} (-1)^{N+1} g(t) + o(N^{-1/3}) \;,
\end{eqnarray} 
 where the function $g(t)$ can be expressed explicitly in terms of $q(t)$ and behaves asymptotically as \cite{moredetails}
 \begin{eqnarray}\label{asympt_g}
 g(t) \sim e^{-\frac{2}{3}t^{3/2}} \; , \; {\rm when} \; t \to \infty \;. 
 \end{eqnarray}
 What happens when one increases $L$ from the critical region 
$L - 2 \sqrt{N} \sim {\cal O}(N^{-1/6})$ towards the large deviation 
regime in the right tail, $L> 2 \sqrt{N}$ (see Fig. \ref{fig_corres}) ? 
As $L$ increases away from $L_c(N)=2 \sqrt{N}$, the amplitude of the second 
term in 
the rhs of Eq. (\ref{beyond}), which is oscillating with $N$, 
increases relatively to the amplitude of the first term. 
And at some value $L \equiv L_{\rm cross}(N)$, it becomes larger than the first 
one: in 
the large deviation regime it becomes the 
leading term (still oscillating with $N$), as 
given in Eq. (\ref{ld}). On the other hand, the first term in the 
rhs of Eq. (\ref{beyond}), $\log{\left({\cal F}_2(t)\right)}$, 
is subdominant for $L > L_{\rm cross}(N)$ and becomes actually the term of 
order ${\cal 
O}(e^{-\frac{4 \pi^2 N}{A} \gamma\left(\frac{A}{\pi^2} \right)})$ 
in Eq. (\ref{in}). Hence, there is a crossover between the two terms in 
the rhs of Eq. (\ref{beyond}) as $L$ crosses the value $L_{\rm cross}(N)$. 
Balancing these
  two terms and making use of the leading behavior of the right tail of 
${\cal F}_2(t)$ together with the asymptotic behavior of 
$g(t)$ given in Eq. (\ref{asympt_g}), one obtains an estimate of $L_{\rm 
cross}(N)$ as~\cite{moredetails}
 \begin{eqnarray}\label{Lc}
 L_{\rm cross}(N) - 2 \sqrt{N} \sim N^{-1/6} \left(\log{N}\right)^{2/3} \; . 
 \end{eqnarray}
 Note that such a crossover is absent in the distribution of the 
largest eigenvalue of GUE random matrices and it is thus a 
specific feature of this vicious walkers problem.  
 
\section{Brownian walkers with absorbing walls: Model II}

In Model II, we have non-intersecting Brownian motions on $[0,L]$ with
absorbing boundary conditions. Comparing \eqref{Fp} and \eqref{ZSpN}
we have the exact identity
\begin{equation}
\tilde F_N(L)= \frac{B_N}{{\hat c}_N}\, e^{\pi^2 N(N+1/2)(N+1))/{6L^2}}\,
L^{-2N^2-N}\,
\mathcal{Z}\left( \frac{2\pi^2 N}{L^2};{\rm Sp}(2N)\right) \;,
\label{F_Z}
\end{equation}
where the constants $B_N$ and ${\hat c}_N$ are independent of $L$.
As in the ${\rm U}(N)$ case, thanks to this identity and known large $N$ 
properties of the partition function $\mathcal{Z}\left( 
A;{\rm Sp}(2N)\right)$ provides us new large $N$ results 
for normalized reunion probability $\tilde F_N(L)$ as a function
$L$. In addition, in this case, $\tilde F_N(L)$ is also identical to
the distribution of the maximal height $H_N$ for non-intersecting 
Brownian excursions. We thus get, as a bonus, new asymptotic results
for the height distribution:
both in the critical regime where its scaling behaviour is described
by the Tracy-Widom distribution ${\cal F}_1(t)$ of GOE matrices, as well
as in its large deviation tails. As in the previous section, below we first 
discuss
the large deviation tails and then the behavior in the critical regime.
 
\subsection{Large deviation tails}\label{s3a}

In this case, using the correspondence $A=2\pi^2 N/L^2$, the critical point 
$A=\pi^2$ in the gauge theory corresponds to a critical value $L=\sqrt{2N}$.
For convenience, we then scale $L=\sqrt{2N} h$ so that the critical
point is now at $h=1$. Choosing $\hat{c}_N = 1$, \eqref{F_Z} then
reads
 \begin{equation}\label{10}
 \tilde{F}_N(  \sqrt{2 N} h) = {B_N \over (\sqrt{2 N} h)^{2N^2 + N}}
 e^{-\pi^2 (N+1/2)(N+1)/12 h^2} \mathcal Z \Big ( {\pi^2 \over h^2}; {\rm Sp} 
(2N) \Big ).
 \end{equation}
 
The partition function (\ref{ZSpN}) can be analyzed in terms of 
the same orthogonal polynomials
appearing in (\ref{RN}). Thus one has \cite[eq.~(23)]{CNS96}
 \begin{equation}\label{23a}
 {d^2 \over d A^2} \log  \mathcal Z(A;{\rm Sp}(2N)) = {1 \over 4 N^2} R_N^{(+)}
 R_{N+1}^{(+)}.
 \end{equation}

Let us first consider the behavior $h>1$ (corresponding to the weak coupling 
phase $A<\pi^2$). 
 Knowledge of the large $N$ form of the polynomials $R_N^{(+)}$ for $A < \pi^2$ allows the
 derivation of the expansion \cite[eq.~(30)]{CNS96}
  \begin{eqnarray}\label{37}
  && \log e^{-A(N+1/2)(N+1)/12}  \mathcal Z(A;{\rm Sp}(2N))  =
    \left(- N^2 - {N \over 2} \right) \log A
   \\
  && +{A \over 8 \pi^2 \sqrt{\pi N}}  \Big ( 1 + {1 \over \sqrt{1 - A/\pi^2}} \Big )
  \Big ( 1 - {A \over \pi^2} \Big )^{-1/4}
  e^{- {2 \pi^2 N \over A}\gamma(A/\pi^2) } + {\cal O}\left(  e^{- {4 \pi^2 N \over A}\gamma(A/\pi^2) } \right) \;, \nonumber
  \end{eqnarray}
  up to terms independent of $A$ (the latter depend on the precise form of $\hat{c}_N$ in
  (\ref{ZSpN})), where $\gamma(x)$ is given by (\ref{gx}).
  
  Substituting (\ref{37}) in (\ref{10}) we see the first term in the former 
cancels,
  allowing us to conclude that for $h > 1$
  \begin{equation}\label{38}
  \tilde{F}_N(  \sqrt{2N} h) \mathop{\sim}\limits_{N \to \infty} 1
  \end{equation}
  and furthermore
   \begin{equation}\label{pdfh}
  \tilde{F}_N'(  \sqrt{2N} h)    \mathop{\sim}\limits_{N \to \infty} 
   e^{- 2 N h^2   \gamma(1/ h^2) } \;,  
   \end{equation}
where $\gamma(x)$ is given in \eqref{gx}. Note that in Model II, the 
derivative $\tilde{F}_N'(L)$ has the interpretation of the probability
density of the maximal height $H_N$ since $\tilde F_N(L)= {\rm Pr}(H_N\le L)$
is the cumulative distribution of height $H_N$. It is useful
now to compare \eqref{pdfh} and
the analogous result \eqref{ld} in Model I. Note that, unlike
in \eqref{ld}, the derivative does not oscillate in sign as $N$ varies.
This is reassuring since in the 
present 
case, $\tilde F_N'(L)$ has the meaning of a probability density 
which is necessarily positive.  
   
We turn now to the large deviation formula for $h < 1$. 
To derive this large deviation formula in this case, in principle
we need to repeat, for $Sp(2N)$, the strong coupling calculation of Douglas 
and 
Kazakov~\cite{DK93} originally done for the ${\rm U}(N)$ case. In practice, 
however, as we show below, one can avoid repeating this calculation
by noting an analogy in the Coloumb gas representation in the two cases
and thereby relating the behavior of $\mathcal Z(A;{\rm Sp}(2N))$
and $\mathcal Z(A;{\rm U}(N))$ in the strong coupling regime.
As a result, we can then directly use the results of Douglas and Kazakov
stated in \eqref{ZA1}.

Let us first recall that for the ratio
of reunion probabilities for Brownian walkers on the circle, we 
relied on the knowledge of
the leading asymptotic form of the Yang-Mills partition function 
$\mathcal Z(A;{\rm U}(N))$ for
$A > \pi^2$ known from \cite{DK93}. The latter in turn is 
deduced using the contrained continuum saddle point
   formula in the Coloumb gas representation
   \begin{eqnarray}\label{ec}
&&  \lim_{N \to \infty} {1 \over N^2} \log e^{A(N^2-1)/24} \mathcal Z(A;{\rm U}(N)) 
\nonumber \\
&& \qquad
= -{A \over 2} \int_{-c}^c x^2 \rho(x) \, dx
   +    \int_{-c}^c  dx   \int_{-c}^c dy \,  \rho(x)  \rho(y) \log |x - y|,
\end{eqnarray}
where the charge density $\rho(x)$ maximizes the rhs subject to the 
constraints
\begin{equation}\label{con}  
   \int_{-c}^c \rho(x) \, dx = 1, \qquad 0 \le \rho(x) \le 1,
\end{equation}
   the latter being a direct signature of the spacing in the lattice gas being $1/N$ (recall the
   discussion above (\ref{ZA1})).
   
In the case of   (\ref{ZSpN}), we begin by supposing all $n_i > 0$, which 
simply alters
$\hat{c}_N$. Then, with $n_i/(2N)$ regarded as the continuous variable,
 the analogue of (\ref{ec}) reads 
 \begin{eqnarray}\label{ec1}
&&  \lim_{N \to \infty} {1 \over N^2} \log  e^{-A(N+1/2)(N+1)/12}  \mathcal 
Z(A;{\rm Sp}(2N)) 
\nonumber \\
&& \qquad
= -A  \int_{0}^d x^2\tilde{\rho}(x) \, dx
   +   \int_{0}^d  dx   \int_{0}^d dy \,  \tilde{\rho}(x)  \tilde{\rho}(y) \log |x^2 - y^2|,
\end{eqnarray}
where the density $\tilde{\rho}(x)$ maximizes the rhs subject to the constraints
\begin{eqnarray}
   \int_{0}^d \tilde{ \rho}(x) \, dx = 1, \qquad 0 \le \tilde{\rho}(x) \le {1 \over 2}.
\end{eqnarray}
   
If we take $\tilde{\rho}(x) =  \tilde{\rho}(-x)$, $\rho(x) = 2 \tilde{\rho}(x)$ and $d=c$
then the rhs of (\ref{ec1}) can be rewritten
\begin{eqnarray}
-A  \int_{-c}^c x^2\tilde{ \rho}(x) \, dx
   +  2 \int_{-c}^c  dx   \int_{-c}^c dy \,  \tilde{\rho}(x) \tilde{\rho}(y) \log |x - y|,
\end{eqnarray}
subject to (\ref{con}).
Since the symmetry of the problem implies $\rho(x)$ is even in (\ref{ec}), this is just twice
the rhs of (\ref{ec}), so we conclude
  \begin{eqnarray}\label{c9}
&&  \lim_{N \to \infty} {1 \over N^2} \log  e^{-A(N+1/2)(N+1)/12}  \mathcal 
Z(A;{\rm Sp}(2N))  \nonumber \\
&& \hspace{2cm} =
2  \lim_{N \to \infty} {1 \over N^2} \log e^{A(N^2-1)/24} \mathcal Z(A;{\rm U}(N)). 
\end{eqnarray}
Using a similar argument an inter-relation of this type was
first noted in \cite{CS95}, although there it is claimed in Eq. (3.7) that $A$ must be replaced by
$A/2$ on the rhs; this in turn is contradicted by the later paper of the same authors
 \cite{CNS96}. We remark that for $A < \pi^2$ (\ref{c9}) is consistent with (\ref{37})
and the appropriate case of (\ref{ZA1}).

For $h <1$, it follows from (\ref{c9}), (\ref{ZA1}) in the case $A > \pi^2$, and (\ref{10}) that
  \begin{equation}\label{fw}
 \tilde{F}_N( \sqrt{2 N} h) \sim e^{-2 N^2 (F_-(1/h^2) - F_+(1/h^2))} \;.
 \end{equation}
 
 \subsection{Double scaling limit}

Having obtained the large $N$ asymptotic behavior of $\tilde F_N(L)$ 
as
a function of $L$
in the left $(L < \sqrt{2N})$ and the right $(L > \sqrt{2N})$ tails, let us 
now focus on the behvaior of $\tilde F_N(L)$
in the vicinity of the critical point, i.e., when $L$ is close
to $\sqrt{2N}$. As in the ${\rm U}(N)$ case, this
corresponds to the double scaling regime near the critical point
$A=\pi^2$ in the ${\rm Sp}(2N)$ gauge theory.
Using results from the gauge theory, we will show here
in a narrow region $|L-\sqrt{2N}|\sim N^{-1/6}$ around
the critical point $L=\sqrt{2N}$, $\tilde F_N(L)$ has the
following scaling behavior
\begin{equation} 
\tilde F_N(L) \to {\cal F}_1\left(2^{11/6} N^{1/6}|L-\sqrt{2N}|\right) \;,  
\label{scaling2}
\end{equation}
where the scaling function ${\cal F}_1(t)$ is precisely the Tracy-Widom
distribution function for GOE random matrices defined in
\eqref{F2-GOE}.

Using results from \cite{DK93}, the double scaling limit of 
(\ref{23a}) has been analyzed in \cite{CNS96}.
In particular, with
\begin{equation}\label{xn}
x_{2N}:= n_c^{2/3} \Big ( 1 - {2N \over n_c} \Big ) \; , \; n_c := {2 N \pi^2 \over A} \;,
\end{equation}
one has from  Eq.~(34) of \cite{CNS96} that in the double scaling limit $N \to \infty$, $x_{2N}$ fixed
\begin{equation}\label{9a}
{d^2 \over d A^2}   \log  \mathcal Z(A;{\rm Sp}(2N))  =
{n_c^4 \over 16 \pi^4 N^2} \Big ( 1 - {2 x_{2N} \over n_c^{2/3}} -
{\pi^4 \over 2 n_c^{3/2}} f_1^2(x_{2N}) - {\pi^2 \over n_c^{2/3}} f_1'(x_{2N}) \Big ).
\end{equation}

To make use of (\ref{9a}), we first note that with the definition
\begin{eqnarray}
W_N(A) :=\log e^{-A(N+1/2)(N+1)/12}  \mathcal Z(A;{\rm Sp}(2N))  \;,
\end{eqnarray}
we have from (\ref{Fp})
\begin{equation}\label{47}
{\partial^2 \over \partial L^2} \log \tilde{F}_N(L) =
{2N^2 + N \over L^2} + {12 N \pi^2 \over L^4} W_N'\Big ( {2N \pi^2 \over M^2} 
\Big ) +
{16 N^2 \pi^4 \over L^6} W_{N}'' \Big ( {2N \pi^2 \over L^2} \Big ).
\end{equation}
Suppose now we fix
\begin{equation}\label{18a}
L^{4/3} \Big ( 1 - {2N \over  L^2} \Big ) := x_{2N} \; .
\end{equation}
For large $N$, upon writing the second factor as the difference of two squares and
setting $L = \sqrt{2N}$ where this does not lead to a zero term, this is equivalent to setting
\begin{equation}\label{18b}
x_{2N} = 2^{7/6} N^{1/6} (L - \sqrt{2N}) \;.
\end{equation}
Recalling (\ref{10}), and with $n_c = ( \sqrt{2N} h)^2$ in (\ref{xn}), it follows from (\ref{9a}), (\ref{47}) and (\ref{18a}) that
\begin{equation}\label{19a}
{\partial^2 \over \partial L^2} \log \tilde{F}_N(L) = - {4 N^2 + 3N \over 
L^2} +
L^2 \Big (1 - 2 {x_{2N} \over L^{4/3}} \Big ) -
L^{2/3} \Big ( {\pi^4 \over 2} f_1^2(x_{2N}) + \pi^2 f_1'(x_{2N}) \Big ) \; .
\end{equation}
But use of
(\ref{18b}) shows that
\begin{eqnarray}
- {4 N^2 + 3N \over L^2} +
L^2 \Big (1 - 2 {x_{2N} \over L^{4/3}} \Big )  = {\cal O}(N^{1/6}) \;, 
\end{eqnarray}
allowing these terms to be ignored in (\ref{19a}), and leaving us with
\begin{equation}\label{19ad}
{\partial^2 \over \partial L^2} \log \tilde{F}_N(L) =  -
L^{2/3} \Big ( {\pi^4 \over 2} f_1^2(x_{2N}) + \pi^2 f_1'(x_{2N}) \Big ).
\end{equation}

Finally,
use (\ref{18b}) to replace the derivative with respect to $L$ on the lhs of (\ref{19ad})
by a derivative with
respect to $x_{2N}$, and substitute for $x_{2N}=x$ according to (\ref{xu}), and $f_1(x)$
for $u(t) = q(t)$ also according to (\ref{xu}). We then see that (\ref{19ad}) implies
\begin{equation}\label{52}
{d^2 \over dt^2} \log  \tilde{F}_N \Big ( \sqrt{2N}(1 + t/(2^{7/3} N^{2/3}) ) \Big) = - {1 \over 2}
\Big ( q^2(t) + q'(t) \Big ).
\end{equation}
Comparison with (\ref{F2}) shows that the distribution of the scaled largest eigenvalue in the GOE
satisfies the same relation, and so we have
\begin{equation}\label{53}
 \tilde{F}_N \Big ( \sqrt{2N}(1 + t/(2^{7/3} N^{2/3}))  \Big )= \mathcal F_1(t).
 \end{equation}
 As in deducing (\ref{33}) from (\ref{32}), to deduce (\ref{53}) from (\ref{52}) we have used also the fact
 that the lhs must tend to 1 as $t \to \infty$, which is a consequence of (\ref{38}).
 
 We saw in the case of the ratio of return probabilities for the walkers on the circle that the
 large deviation formula relating the values smaller than the mean is connected to the left
 tail of the double scaling distribution about the mean. The present problem of the cumulative distribution for the maximum displacement of non-intersecting Brownian walkers near a wall exhibits the same feature. Thus making use of the expansion (\ref{34}) in (\ref{fw}) shows
\begin{eqnarray}
\tilde{F}_N \Big ( \sqrt{2N}(1 + t/(2^{7/3} N^{2/3}) ) \Big ) \mathop{\sim}\limits_{t \to -\infty}
e^{ {t^3 \over 24}} \;,
\end{eqnarray}
which is indeed the leading order tail form of $\mathcal{F}_1(t)$ \cite{TW96,RRV06}. On the other hand, the right tail of the distribution $\tilde F_N(L)$ for $L > \sqrt{2N}$ exhibits a crossover exactly similar to the one described above for non-intersecting Brownian motions on a circle (\ref{beyond}, \ref{Lc}) \cite{moredetails}. Note however that in that case, the second term as in Eq. (\ref{beyond}) does not oscillate with $N$ because in that case $\tilde F_N(L)$ has an interpretation in terms of cumulative distribution.   

The result (\ref{53}) is in keeping with known results about fluctuating interfaces belonging to the universality class of 
the Kardar-Parisi-Zhang (KPZ) equation in $1+1$ dimensions. Indeed, the top path of such watermelons configuration (see Fig. \ref{fig:watermelon}) can be mapped, in the limit $N \to \infty$, onto the height field of such KPZ interface in the so-called ''droplet'' ({\it i.e.} curved) geometry \cite{PS99}. The extreme value statistics of such interface in the KPZ universality class and in curved geometry has recently attracted some attention \cite{Jo03, RS10}. In particular, using the fact that the maximal value of the height field in the droplet
geometry can be mapped onto the height field (at a given point) in the flat geometry \cite{KMH92} it was shown, albeit indirectly in Ref. \cite{Jo03}, that
the distribution of $H_N$ (see Fig. \ref{fig:watermelon}), correctly shifted and scaled, is indeed described by ${\cal F}_1(t)$. Here, we obtain this result by a direct computation of the distribution of $H_N$ in the large $N$ limit. Moreover the $\mathcal F_1(t)$ fluctuations have
previously been established in the case of the distribution of the displacement $M$ of the right-most 
walker amongst $N$ returning vicious walkers in discrete time and
in the presence of a wall,
and with the technical requirement that twice the number of walkers be greater than the total number of steps
\cite{BF00}. This  model gives rise the matrix integral
\begin{equation}\label{Sp}
\Big \langle {\rm Tr} S^{2N}  \Big \rangle_{S \in {\rm Sp}(2M)} \;,
\end{equation}
which upon Poissonization in $N$ is the symplectic analogue of (\ref{firsteqn}).
In fact the Poissonized form of (\ref{Sp}) appears in Hammersley model of directed last passage
percolation as revised in
Section 1.1, but with the points confined to be below the line $y=x$ in the square
\cite{BR01a}, \cite[\S 10.7.1]{Fo10}.

\section{Conclusion}

To conclude, we have studied the normalized reunion probability of $N$ non\hyph{intersecting} Brownian
motions confined on a line segment $[0,L]$ with three different types of boundary conditions : (I) periodic (where
the Brownian walkers are thus moving on a circle of radius $L/2 \pi$) (\ref{28}), (II) absorbing boundary conditions at both extremities $0$ and $L$ (\ref{RL}) and
(III) reflecting boundary conditions at both ends (\ref{RLIII}). We have shown that, in each of these models, 
this quantity is given (up to a prefactor that we have computed) by the partition function of $2-d$ Yang-Mills theory on
the sphere with a given gauge group, and computed with the heat-kernel action. We have found that models I, II and III correspond respectively to the group 
${\rm U}(N)$ (\ref{corrUN}), ${\rm Sp}(2N)$ (\ref{corrSP2N}) and ${\rm SO}(2N)$ (\ref{corrSO2N}). Borrowing results from these different field theories, we have shown that, in the large $N$ limit, these reunion 
probabilities exhibit a third-order phase transition as $L$ crosses a critical value $L_c \sim \sqrt{N}$. The region corresponding to $L > L_c$, which corresponds
to the weak coupling regime in the Yang-Mills theory, describes the right tail of this normalized reunion probability, while the region $L < L_c$, which corresponds
to the strong coupling regime, describes its left tail (Fig. \ref{fig_corres}). In the critical region of width $N^{-1/6}$, close to $L_c$, 
one finds that the reunion probability, correctly shifted and scaled, converges to the Tracy-Widom distribution corresponding respectively to GUE, ${\cal F}_2(t)$, in model I and
to GOE, ${\cal F}_1(t)$, in model II and III. One of the 
main achievements of this paper to relate the Painlev\'e equation which 
describes the singularity of the free
energy in the double scaling limit of these $2-d$ Yang-Mills theories with the one defining the Tracy-Widom distributions. In the case of model II, the normalized
reunion probability has the interpretation of the cumulative distribution of the maximal height of the corresponding watermelon configuration (Fig. \ref{fig:watermelon}). 
Our results thus show directly that this cumulative distribution is given in the large $N$ limit by ${\cal F}_1(t)$, a result was obtained before in a rather indirect 
way in Ref. \cite{Jo03}.          

In this paper, we have thus presented the correspondence between boundary conditions (in the
vicious walkers models) and gauge groups (in Yang-Mills theories in two dimensions on a sphere) and discussed its
consequences but the deep reason behind it deserves certainly further study. An alternative way to explore these connections
could be to study the relations between vicious walkers models and Chern-Simons theory as in Ref. \cite{H05, HT04, ST10}. Yet another
point of view could be to adopt the formulation of these vicious walkers problems in terms of Dyson's Brownian motion. Indeed, it can be shown that the propagator 
of this process, expressed as a path integral, is precisely given by the partition function of Yang-Mills theory on the sphere 
with the appropriate gauge group, depending on the boundary conditions \cite{MP94}. We hope that this will stimulate further works along these directions.

\section*{Acknowledgments}
We thank Alain Comtet for many useful discussions and a careful reading
of the manuscript. 
PJF thanks Bernard Jancovici for the opportunity to again visit the Laboratoire de Physique Th\'eorique
at Orsay, and so make this collaboration possible, and also acknowledges the support of the
Australian Research Council.

\begin{appendix}

\section{Ratio of reunion probabilities for non\hyph{intersecting} 
Brownian motions on a circle}\label{appendix}

In this appendix, we derive the formula given in Eq. (\ref{29}) 
for the ratio $\tilde G_N(L) = {\tilde R_L^I(1)}/{\tilde R_\infty^I(1)}$ 
introduced in Eq. (\ref{28}). 
We consider $N$ non-intersecting Brownian motions on a line segment $[0,L]$
with periodic boundary conditions or equivalently
on a circle of radius $L/2 \pi$,
starting in the vicinity of the origin at time $\tau = 0$.
The reunion probability $\tilde R_L^I(1)$ denotes the probability that 
the walkers return to their initial configuration at time $\tau = 1$.

Let us begin with the case of $N$ free Brownian motions, 
with a diffusion constant $D=1/2$, on a circle of radius 
$L/2\pi$. The position of the $N$ walkers are thus 
labelled by the angles $\theta_1, \cdots, \theta_N$. We denote by $P_N(\theta_1, \cdots, \theta_N; t | \rho_1, \cdots, \rho_N;0)$ the probability that the positions of the $N$ walkers are $\theta_1, \cdots, \theta_N$ at time $t$, given that their positions were $\rho_1, \cdots, \rho_N$ at initial time. It is easy to see that $P_N(\theta_1, \cdots, \theta_N; t | \rho_1, \cdots, \rho_N;0)$ satisfies the Fokker-Planck equation
\begin{eqnarray}\label{FP}
&&\frac{\partial}{\partial t} P_N = \frac{2 \pi^2}{L^2} \sum_{k=1}^N \frac{\partial^2}{\partial \theta_k^2} P_N \;, \nonumber \\
&& P_N(\theta_1, \cdots, \theta_N; t=0 | \rho_1, \cdots, \rho_N;0) = \prod_{k=1}^N \delta(\theta_k - \rho_k) \;,
\end{eqnarray}
which simply comes from the expression of the bi-dimensional Laplacian in terms of polar variables (we recall that the radius of the circle is $L/2\pi$ and the diffusion coefficient is $D=1/2$), together with the constraint of periodicity
\begin{eqnarray}\label{periodicity}
P_N(\cdots, \theta_k + 2\pi, \cdots; t | \rho_1, \cdots, \rho_N;0) &=& P_N(\cdots, \theta_k + 2\pi, \cdots; t | \rho_1, \cdots, \rho_N;0) \;, \nonumber \\
&&\; \forall \; 1 \leq k \leq N \;,
\end{eqnarray}
and similarly for a shift of $2 \pi$ of the variables $\rho_k$'s. Therefore, from Eq. (\ref{FP}) and Eq. (\ref{periodicity}), $P_N$ can be written as the propagator, in imaginary time, of $N$ independent quantum free particles on a circle of circumference $L$. Using path integral techniques, one thus writes (using the notation ${\boldsymbol \theta} \equiv (\theta_1, \cdots, \theta_N)$)
\begin{eqnarray}
P_N(\theta_1, \cdots, \theta_N; t | \rho_1, \cdots, \rho_N;0) = \langle {\boldsymbol \theta} | e^{- t \hat H_L} | \boldsymbol \rho \rangle \;,
\end{eqnarray}
with $\hat H_L = \sum_{i=k}^N \hat h_{L,k}$ where $\hat h_{L,k} = -\frac{2 \pi^2}{L^2} \frac{\partial^2}{\partial \theta_k^2}$ has to be understood as the Hamiltonian of a free particle on a circle of circumference $L$ so that the allowed eigenvalues of $\hat h_{L,k}$ are $E_{n_k} = \frac{2\pi^2 n_k^2}{L^2}$ associated to the eigenvectors $\phi_{n_k}(\theta) = \frac{1}{\sqrt{2 \pi}} e^{i n_k \theta}$, $n_k \in \mathbb{N}$. Therefore one has in that case
\begin{eqnarray}\label{path_integral}
P_N(\theta_1, \cdots, \theta_N; t | \rho_1, \cdots, \rho_N;0)  = \sum_{E} \Psi_{E}(\theta_1, \cdots, \theta_N) \Psi^*_{E}(\rho_1, \cdots, \rho_N) e^{-E t} \;,
\end{eqnarray} 
with $E = E_{n_1} + \cdots + E_{n_N}$ and $\Psi_{E}(\theta_1, \cdots, \theta_N) = \langle {\boldsymbol \theta} | E \rangle $ is the manybody eigenfunction of $\hat H_L$. For independent Brownian motions (without the non-crossing condition), $\Psi_E(\theta_1, \cdots, \theta_N)$ is simply the product of the single particle wave function $\Psi_E(\theta_1, \cdots, \theta_N) =  \prod_{i=1}^N \phi_{n_i}(\theta_i)$. 

We can now consider the problem of $N$ non-intersecting Brownian motions on a circle of circumference $L$ and study the corresponding propagator $P_N(\boldsymbol \theta; t | \boldsymbol \rho;0)$. It satisfies the same equations as before (\ref{FP}, \ref{periodicity}) together with the additional non-crossing constraint:
\begin{eqnarray}\label{noncolliding}
P_N(\theta_1, \cdots, \theta_N; t | \rho_1, \cdots, \rho_N;0) = 0 \; {\rm if} \; \theta_j = \theta_k \; {\rm for \, any \, pair \,} j,k \;. 
\end{eqnarray}
Following Ref. \cite{SMCR08,NM09}, this propagator can be computed using the 
path-integral formalism explained above (\ref{path_integral}) where, to incorporate the non-colliding condition, the many-body wave function $\Psi_E(\theta_1, \cdots, \theta_N) $ must be Fermionic, {\it i.e.} it vanishes if any of the two coordinates are equal. This anti-symmetric wave function is thus constructed from the one particle wave functions $\phi_{n_i}$ of $\hat h_i$ by forming the associated Slater determinant.  Therefore one has, in that case
\begin{eqnarray}\label{slater}
\psi_E(\boldsymbol \theta) = \frac{1}{\sqrt{N!}} \det_{1 \leq j,k \leq N} \phi_{n_j}(\theta_k) \;, \; E = \frac{2\pi^2}{L^2} \sum_{k=1}^N n_k^2 \;.
\end{eqnarray} 

From the propagator $P_N$, we can now compute the reunion probability, which is the probability that all the $N$ walkers start and end at the same position on the circle, say $\theta_1 = \theta_2 = \cdots = \theta_N = 0$, on the unit time interval. However, such a probability is ill defined for a system in continuous space and time. We can go around this problem by assuming that the starting and finishing positions (angles) of the $N$ walkers are $0 < \epsilon_1 < \epsilon_2 < \cdots < \epsilon_N$ and only at the end take the limit $\epsilon_i \to 0$. Therefore we can compute the ratio of reunion probabilities $\tilde G_N(L)$ as
\begin{eqnarray}\label{startexpr_app}
 \tilde G_N(L) = \lim_{\epsilon_i \to 0} \frac{\langle \boldsymbol \epsilon | e^{- \hat H_L} | \boldsymbol \epsilon\rangle}{\langle \boldsymbol \epsilon | e^{- \hat H_\infty} | \boldsymbol \epsilon\rangle} \;,
\end{eqnarray}
where $\hat H_\infty$ denotes the Hamiltonian of the $N$ walkers on the full real axis. Using the expressions in Eqs (\ref{path_integral}, \ref{slater}), one checks that, in the limit $\epsilon_i \to 0$, powers of $\epsilon_i$'s cancel between the numerator and the denominator in Eq. (\ref{startexpr_app}) yielding the expression for $\tilde G_N(L) $ given in Eq. (\ref{29}) in the text. 

For instance, for $N=1$ one has
\begin{eqnarray}
 \tilde G_1(L) = \frac{\sqrt{2 \pi}}{L} \sum_{n = -\infty}^\infty e^{-\frac{2 \pi^2}{L^2} n^2} \;,
\end{eqnarray}
which can also be written, using the Poisson summation formula
\begin{eqnarray}
 \tilde G_1(L) = \sum_{n = -\infty}^\infty e^{- \frac{L^2}{2} n^2} \;.
\end{eqnarray}
In particular, we obtain the large $L$ behavior as
\begin{eqnarray}\label{direct_N1}
 \tilde G_1(L) = 1 + 2 e^{- \frac{L^2}{2}} + {\cal O}(e^{-2 L^2}) \;,
\end{eqnarray}
showing explicitly that $ \tilde G_1(L)$ does not have the meaning of a cumulative distribution. 

\section{Reunion probability with absorbing and reflecting boundary conditions}
Here we briefly outline the derivations of the results for ${\tilde F}_N(L)$
in \eqref{Fp} and ${\tilde E}_N(L)$ in \eqref{ENL}. In the first case, we
again have $N$ non-intersecting Brownian motions on the line segment $[0,L]$
with absorbing boundary conditions at $0$ and $L$. The walkers all start at 
time $\tau=0$ in the vicinity of the origin and we want to compute their reunion probability $R_L^{II}(1)$ near the origin after time $\tau=1$.
The calculation proceeds in the same way as in the periodic case in the 
previous appendix. One writes the Fokker-Planck equation for the
probability density $P_N(x_1,x_2,\ldots, x_N;t|y_1,\ldots, y_N;0)$   
of reaching $\{x_1,x_2,\ldots, x_N\}$ at time $t$ starting from the
initial positions $\{y_1,y_2,\ldots,y_N\}$. This reads
\begin{eqnarray}\label{FP2}
&&\frac{\partial}{\partial t} P_N = 
\frac{1}{2} \sum_{k=1}^N 
\frac{\partial^2}{\partial x_k^2} P_N \;, \nonumber \\
&& P_N(x_1, \cdots, x_N; t=0 | y_1, \cdots, y_N;0) 
= \prod_{k=1}^N \delta(x_k - y_k).
\end{eqnarray}
One then writes the solution using path integrals exactly as in 
\eqref{path_integral}.
The rest of the calculation is similar as in Appendix A. The only 
difference is that in constructing the Slater determinant, one now
has to use the normalized single particle wave function as 
\begin{equation}
\phi_{n_k}(x)= \sqrt{\frac{2}{L}}\, \sin\left(\frac{n_k \pi x}{L}\right) \;,
\label{spwf-abs}
\end{equation}
which satisfies the absorbing boundary condition $\phi_{n_k}(x)=0$ at
$x=0$ and $x=L$. Using this, one just repeats the calculation
of Appendix A to derive the result for ${\tilde F}_N(L)$
in \eqref{Fp}.

In the reflecting case, the only change is again in the normalized 
single particle wave function that reads
\begin{equation}
\phi_{n_k}(x)=\sqrt{\frac{2}{L}}\, \cos\left(\frac{n_k \pi x}{L}\right)
\label{spwf-ref}
\end{equation}
and satisfies the reflecting boundary condition $\partial_x \phi_{n_k}(x)=0$
at $x=0$ and $x=L$. Repeating the rest of the calculation with this 
modification as in Appendix A, one easily derives the result for 
${\tilde E}_N(L)$ in \eqref{ENL}.

\end{appendix}


\providecommand{\bysame}{\leavevmode\hbox to3em{\hrulefill}\thinspace}
\providecommand{\MR}{\relax\ifhmode\unskip\space\fi MR }
\providecommand{\MRhref}[2]{%
  \href{http://www.ams.org/mathscinet-getitem?mr=#1}{#2}
}
\providecommand{\href}[2]{#2}

\end{document}